\documentclass[twocolumn,article,notitlepage,floatfix,longbibliography,nofootinbib]{revtex4-1}

\usepackage{amsmath}
\usepackage{dsfont}
\usepackage{verbatim}
\usepackage{braket}
\usepackage{titlesec}
\usepackage{graphicx}
\usepackage[usestackEOL]{stackengine}
\usepackage{array}
\usepackage{tikz}
\usepackage{xcolor}
\usepackage{epsfig} 
\usepackage{amssymb} 
\usepackage{longtable} 
\usepackage{float}
\usepackage{mathtools}
\usepackage{xparse}
\usepackage{times}
\usepackage[normalem]{ulem}

\renewcommand{\eqref}[1]{Eq. ({\ref{#1}})}

\begin{document}

\title{Regularized lattice theory for spatially dispersive nonlinear optical conductivities}

\author{Steven Gassner}
\affiliation{\\\,\\
Department of Physics and Astronomy, University of Pennsylvania, Philadelphia, PA 19104}
\author{E. J. Mele}
\affiliation{\\\,\\
Department of Physics and Astronomy, University of Pennsylvania, Philadelphia, PA 19104}

\date{\today}

\begin{abstract}
Nonlinear optical responses are becoming increasingly relevant for characterizing the symmetries and quantum geometry of electronic phases in materials. Here, we develop an expanded diagrammatic scheme for calculating \textit{spatially dispersive corrections} to nonlinear optical conductivities, which we expect to enhance or even dominate even-order responses in materials of recent interest. Building upon previous work that enforces gauge invariance of spatially uniform nonlinear optical responses, we review the cancellation of diagrams required to ensure the equivalence between velocity gauge and length gauge formulations, and provide a simple vertex rule for extending optical responses to first order in the light wave vector $\mathbf{q}$. We then demonstrate the method with calculations on a prototypical centrosymmetric model where spatial dispersion admits anomalous second-harmonic generation, a response that is symmetry-forbidden under the dipole approximation.
\end{abstract}

\maketitle

\section{Introduction}

Optical responses are an extremely valuable tool for understanding the electronic structure of quantum matter, from materials to molecules \cite{Boyd2003,Sekino1986}. Not only are they vital for characterizing a host of different devices, from photovoltaics to lasers, but they are also immensely useful as probes for symmetry breaking and quantum geometric properties in materials \cite{Young2012, Morimoto2016, Wu2017, deJuan2017, Ma2017, Chan2017, Holder2020, Ni2022, Ahn2022b, Tai2023}. Over the past several decades, the development of a straightforward and general means of systematically calculating these responses has been a surprisingly difficult task, revealing a number of subtleties that are still being worked out to this day. A main reason is that there are different choices of gauge for describing the light-matter coupling with various advantages and disadvantages. One option is to couple the electric field $\textbf{E}(t)$ with the position operator $\hat{\textbf{r}}$,
\begin{equation}
    \hat{\mathcal{H}} \to \hat{\mathcal{H}} - {\rm q}\, \hat{\textbf{r}}\cdot \textbf{E}(t)
    \text{,}
\end{equation}
which goes by the name ``length gauge" (also known as ``position gauge" or "dipole gauge.") Here, ${\rm q} = -|e|$ denotes the electron charge. Another option, called the ``velocity gauge," employs a minimal substitution in the $\mathbf{k}$-dependent Bloch Hamiltonian,
\begin{equation}
    \hat{\mathcal{H}}(\mathbf{k}) \to \hat{\mathcal{H}}(\mathbf{k} - {\rm q}\,\mathbf{A}(t))
    \text{,}
\end{equation}
where $\mathbf{A}(t)$ is the magnetic vector potential satisfying $\mathbf{E}(t) = -\partial_t\mathbf{A}(t)$. (At first order in $\mathbf{A}(t)$, this takes the form of a coupling $-{\rm q}\,\hat{\mathbf{v}}\cdot\mathbf{A}(t)$ where $\hat{\mathbf{v}} = \nabla_\mathbf{k}\hat{\mathcal{H}}$ is the velocity operator; hence the name ``velocity gauge.") These two gauges are related by a time-dependent unitary transformation, and are hence equivalent \cite{Kobe1978} in a fully microscopic theory. However maintaining this equivalence for effective Hamiltonians that are projected into a partial sector of the full Hilbert space requires additional consideration.

Velocity gauge, while having the advantage of being diagonal in \textbf{k}-space and easier for numerical calculations, has the serious disadvantage of introducing artificial low-frequency divergences that must be regularized systematically. Historically, this has been attempted using sum rules, first developed by Aversa and Sipe \cite{Sipe1993, Aversa1995} and later generalized by Ventura et al \cite{Ventura2017, Passos2018}, which formally show that the weight of these divergent terms vanishes only after a full $\textbf{k}$-space integration. In more recent work with Wannierized tight-binding models, Sch\"uler et al identify a sum rule encoding the cancellation between the paramagnetic and diamagnetic currents  calculated at {\it linear order} in $\mathbf{A}(t)$, and show that enforcing this sum rule improves the accuracy of velocity gauge calculations \cite{Schuler2021}.

In this work, we relate a convenient diagrammatic method \cite{Parker2019} and the traditional reduced density matrix perturbation theory approach \cite{Ventura2017, Passos2018}. Representing terms in the perturbation theory by their associated diagrams, we can understand in a physically transparent manner the general cancellation between the paramagnetic and diamagnetic contributions in the linear response, and investigate the analogs of that cancellation that occurs for nonlinear responses. We also introduce a scheme for extending these results to first order in the wavevector $\mathbf{q}$ of the incident radiation. We call these responses \textit{spatially dispersive}, because they take into account at lowest order the effects of the spatial gradient of the electric field on the optical response. As a prototypical example, we investigate the anomalous second-harmonic generation (SHG) response that appears in a centrosymmetric 1D model. While SHG is typically understood to vanish in centrosymmetric systems in the dipole approximation (and hence is often used experimentally as a probe for inversion breaking in crystals) this selection rule is violated when one takes into account the spatial variation of the electromagnetic field on the length scale of the electronic states being coupled in optical transitions. This becomes especially relevant for artificial lattices where the lattice constants are inflated and can greatly exceed the microscopic atomic scale or more generally for band structures in which the Wannier representation introduces coherence over large length scales.

\section{Calculations of nonlinear optical responses}

\subsection{Generalities}

To calculate nonlinear optical conductivities, we employ the density matrix equation of motion approach detailed in Refs. \cite{Ventura2017, Passos2018} and connect it to the diagrammatic approach developed in Ref. \cite{Parker2019}. Recall the density matrix equation of motion,
\begin{equation}
    i\hslash\,\partial_t \hat{\rho}(t) = [\hat{\mathcal{H}}(t),\hat{\rho}(t)]
    \text{,}
\end{equation}
Assume that $\hat{\mathcal{H}}(t) = \hat{\mathcal{H}}^{(0)} + \hat{V}(t)$, with $\hat{V}(t)$ a weak perturbation. Transforming to Fourier representation,
\begin{equation}
    \hat{\rho}(t) = \int d\omega\,e^{-i\omega t} \, \hat{\rho}(\omega)
    \text{,}
\end{equation}
\begin{equation}
    \hat{V}(t) = \int d\omega\,e^{-i\omega t} \, \hat{V}(\omega)
    \text{,}
\end{equation}
we can re-express the equation of motion as an iterative equation in the eigenbasis of the unperturbed Hamiltonian, $\hat{\mathcal{H}}^{(0)}_{ab} = \epsilon_a \delta_{ab}$,
\begin{equation*}
    (\hslash\omega^{(n)} - \epsilon_{ab})\rho^{(n)}_{ab}(\omega^{(n)}) = [\hat{V}(\omega_n),\hat{\rho}^{(n-1)}(\omega^{(n-1)})]_{ab}
\end{equation*}
\begin{equation}
    \implies \rho^{(n)}_{ab}(\omega^{(n)}) = \frac{[\hat{V}(\omega_n),\hat{\rho}^{(n-1)}(\omega^{(n-1)})]_{ab}}{\hslash\omega^{(n)} - \epsilon_{ab}}
    \text{,}
\end{equation}
with $\omega^{(n)} = \sum_{j=1}^n \omega_j$ and $\epsilon_{ab} \equiv \epsilon_a - \epsilon_b$, where $\epsilon_a$ are the eigenvalues of $\hat{\mathcal{H}}^{(0)}$. Introducing a matrix $\hat{\epsilon}$ whose entries are $\epsilon_{ab}$, (and henceforth setting $\hslash=1$), we can compactly write the solution to the equation of motion as,
\begin{equation}
    \hat{\rho}^{(n)}(\omega^{(n)}) = \frac{1}{\omega^{(n)}-\hat{\epsilon}} \circ \left[\hat{V}(\omega_n) \,,\, \hat{\rho}^{(n-1)}(\omega^{(n-1)})\right]
    \text{,} \label{EOM}
\end{equation}
where $\circ$ denotes the Hadamard product, or elementwise multiplication: $(\hat{A}\circ\hat{B})_{ab} = A_{ab}B_{ab}$. As a final bit of notation, we will assume $\hat{V}$ is a sum of terms from which we can select a different term $\hat{V}_i$ with each iteration $i$ of Eq. (\ref{EOM}). We therefore write,
\begin{equation}
    \hat{\rho}^{(n)}_{\hat{V}_1\hdots\hat{V}_n} = \frac{1}{\omega^{(n)}-\hat{\epsilon}} \circ \left[\hat{V}_n \,,\, \hat{\rho}^{(n-1)}_{\hat{V}_1\hdots\hat{V}_{n-1}}\right]
    \text{,}\label{notation}
\end{equation}
dropping the implied dependence of $\hat{\rho}^{(n)}$ on $\omega^{(n)}$ as well as the dependence of $\hat{V}_n$ on $\omega_n$. The utility of this notation will become clear shortly.

\subsubsection{Velocity gauge: Diagrammatic method}

Let us first apply this density matrix perturbation theory to light-matter coupling in velocity gauge. In this case,
\begin{equation}
    \hat{\mathcal{H}}_\mathbf{k}(t) = \hat{\mathcal{H}}^{(0)}_{\mathbf{k}+e\mathbf{A}(t)} \equiv \hat{\mathcal{H}}^{(0)}_\mathbf{k} + \hat{V}_\mathbf{k}(t)
    \text{,}
\end{equation}
which means that $\hat{V}_\mathbf{k}(t)$ takes the form,
\begin{equation}
\begin{split}
    \hat{V}_\mathbf{k}(t) &= \frac{\partial\hat{\mathcal{H}}_\mathbf{k}^{(0)}}{\partial k_{\alpha_1}} e A_{\alpha_1}(t) + \frac{1}{2} \frac{\partial^2\hat{\mathcal{H}}_\mathbf{k}^{(0)}}{\partial k_{\alpha_1}\partial k_{\alpha_2}} e^2 A_{\alpha_1}(t)A_{\alpha_2}(t) + \hdots \\
    &= \sum_{n=1}^\infty \frac{e^n}{n!}\hat{h}^{\alpha_1\hdots\alpha_n}A_{\alpha_1}(t)\hdots A_{\alpha_n}(t)
    \text{,}
\end{split}
\end{equation}
where we introduce the shorthand $\hat{h}^{\alpha_1\hdots\alpha_n} \equiv \partial^{\alpha_1}\hdots\partial^{\alpha_n}\hat{\mathcal{H}}^{(0)}_\mathbf{k}$ with $\partial^\alpha \equiv \frac{\partial}{\partial k_\alpha}$. Here and throughout, Greek indices denote spatial directions, and repeated indices are summed over. We can calculate the density matrix to any desired order in $A_\alpha(t)$. At first order, we have,
\begin{equation}
    \hat{\rho}^{(1)}(\omega_1) = \frac{1}{\omega_1-\hat{\epsilon}}\circ\left[\hat{h}^{\alpha_1},\hat{\rho}^{(0)}\right] A_{\alpha_1}(\omega_1)
    \text{.}
\end{equation}
At second order, we have an additional term,
\begin{equation}
\begin{split}
    \hat{\rho}^{(2)}(\omega^{(2)}) &= \frac{1}{\omega^{(2)}-\hat{\epsilon}}\circ\left[\hat{h}^{\alpha_2},\hat{\rho}^{(1)}(\omega_1)\right]A_{\alpha_2}(\omega_2) \\
    &\,\,\,\,\,\,\,+ \frac{1}{\omega^{(2)}-\hat{\epsilon}}\circ\left[\frac{1}{2}\hat{h}^{\alpha_1\alpha_2},\hat{\rho}^{(0)}\right]A_{\alpha_1}(\omega_1)A_{\alpha_2}(\omega_2) \\
    &\equiv \left(\hat{\rho}^{(2)}_{\hat{h}^{\alpha_1},\hat{h}^{\alpha_2}}(\omega^{(2)}) + \hat{\rho}^{(1)}_{\frac{1}{2}\hat{h}^{\alpha_1\alpha_2}}(\omega^{(2)})\right)A_{\alpha_1}(\omega_1)A_{\alpha_2}(\omega_2)
    \text{.}
\end{split}
\end{equation}
Hence, the utility of the notation in Eq. (\ref{notation}) is to split the density matrix at $n$-th order in $A_\alpha(t)$ into a sum of density matrices $\hat{\rho}^{(m)}_{\hat{V}_1\hdots\hat{V}_m}$ that are perturbed with relatively simple operators $\hat{V}_i$. In fact, each of these operators $\hat{V}_i$ is simply a $j$-th $\mathbf{k}$-derivative (for some $j$) of the bare Hamiltonian $\hat{\mathcal{H}}^{(0)}_\mathbf{k}$, which has the physically transparent meaning of a vertex interaction with $j$ photons. (A word of caution: the ``$(n)$" in $\hat{\rho}^{(n)}_{\hat{V}_1\hdots\hat{V}_n}$ no longer means ``at $n$-th order in $A_\alpha(t)$" (as opposed to the more conventional notation $\hat{\rho}^{(n)}$) and now simply means ``perturbed with $n$ operators $\hat{V}_i$, $i=1,\hdots,n$.")

\begin{figure}
    \centering
    \includegraphics[width=0.7\linewidth]{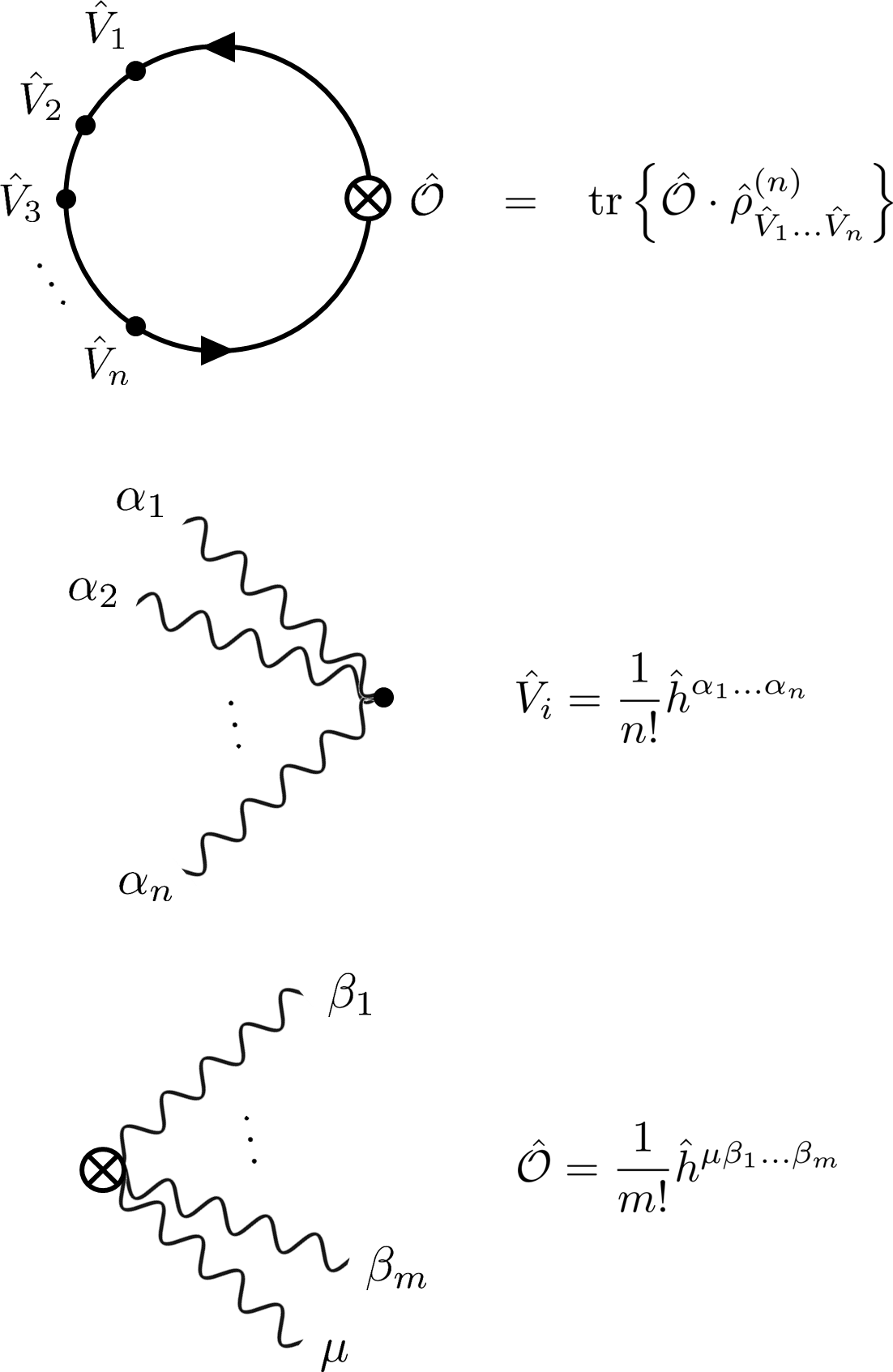}
    \caption{Summary of Feynman rules employed for computing optical responses in velocity gauge. Following the convention of \cite{Parker2019}, the output vertex is marked with a ``$\otimes$," and represents some output current operator $\hat{\mathcal{O}}$. Measuring this output current amounts to a trace of this operator with a density matrix, which is represented by a fermion loop. This density matrix is perturbed (as in Eq. (\ref{notation})) with operators $\hat{V}_i$ represented by black vertices. The vertex operators take the form of $\mathbf{k}$-derivatives of the Hamiltonian (denoted $\hat{h}^{\alpha_1\hdots\alpha_n}$) with the directions of the derivatives given by the polarizations (denoted with Greek letters) of the photons entering the vertex. The multi-photon vertex operators are symmetrized with a factor $1/n!$, where $n$ is the number of photons \textit{excluding} the photon representing the output current (we reserve the index $\mu$ to indicate the polarization of the output photon).}
    \label{fig:feynman rules}
\end{figure}

This decomposition is the basis for the connection with the diagrammatic scheme in Ref. \cite{Parker2019}. At zeroth order in $A_\alpha(t)$, the expectation value of the current is simply a trace of the current operator with respect to the equilibrium density matrix,
\begin{equation}
    \left\langle\hat{j}^\mu\right\rangle_0 = \text{tr}\left\{\hat{j}_{(0)}^\mu\cdot\hat{\rho}^{(0)}\right\}
    \text{.}
\end{equation}
At nonzero order in $A_\alpha(t)$, both the density matrix \textit{and} the current operator must be expanded in powers of $A_\alpha(t)$,
\begin{equation}
\begin{split}
    \hat{j}^{\mu}(t) &\equiv -\frac{\partial\hat{\mathcal{H}}_{\mathbf{k}}(t)}{\partial A_\mu(t)} \\
    &= -\sum_{n=0}^\infty \frac{e^{n+1}}{n!}\hat{h}^{\mu\alpha_1\hdots\alpha_n}A_{\alpha_1}(t)\hdots A_{\alpha_n}(t)
    \text{,} \label{j expansion}
\end{split}
\end{equation}
\begin{equation}
    \hat{\rho}(t) = \sum_{n=0}^{\infty} \hat{\rho}^{(n)}(t)
    \text{.}
\end{equation}
Keeping only terms at some desired order in $A_\alpha(t)$, the result is always a sum of traces, where each trace (expressible by a diagram) is acting on a product between some term in Eq. (\ref{j expansion}) and a density matrix of the form $\hat{\rho}^{(n)}_{\hat{V}_1\hdots\hat{V}_n}$. We express this diagrammatically in terms of ``Feynman rules" illustrated in Figure \ref{fig:feynman rules}. This approach can be viewed as a hybrid approach making connection between the density matrix equation of motion method of Refs. \cite{Ventura2017, Passos2018} and the Matsubara Green's function approach of Ref. \cite{Parker2019}, which lead to the same results. That is, one can alternatively view the calculation of $\hat{\rho}^{(n)}_{\hat{V}_1\cdots\hat{V}_n}$ in a particular diagram using a trace of a product of Matsubara Green's functions multiplied by vertex operators, summed over a single unconstrained fermionic Matsubara frequency.

The last step is to convert the response in terms of $A_{\alpha_j}(\omega_j)$ to a response in terms of $E_{\alpha_j}(\omega_j) \equiv i\omega_j A_{\alpha_j}(\omega_j)$. We formally define,
\begin{equation}
\begin{split}
    \left\langle j^\mu(\omega^{(n)}) \right\rangle =\,\,& \sigma^{\mu\alpha_1\hdots\alpha_n}(\omega^{(n)};\omega_1,\hdots,\omega_n) \\
    &\times E_{\alpha_1}(\omega_1)\hdots E_{\alpha_n}(\omega_n)
    \text{.}
\end{split}
\end{equation}
Recalling $\omega^{(n)}\equiv\sum_{j=1}^{n} \omega_j$, this is the fully general nonlinear conductivity at frequency $\omega^{(n)}$ due to input electric fields at frequencies $\{\omega_j\}_{j=1,\hdots,n}$. We calculate this by first calculating a quantity which we will call $\kappa^{\mu\alpha_1\hdots\alpha_n}_\mathbf{k}(\omega^{(n)};\{\omega_j\})$
\begin{equation}
\begin{split}
    \sigma^{\mu\alpha_1\hdots\alpha_n}(\omega^{(n)};\{\omega_j\}) =\,\,& -\frac{e^{n+1}}{\hslash^n}\prod_{j=1}^{n}\left(\frac{1}{i\omega_j}\right)\\
    &\times\int[d\mathbf{k}]\,\kappa_\mathbf{k}^{\mu\alpha_1\hdots\alpha_n}(\omega^{(n)};\{\omega_j\})
    \text{.}
\end{split}
\end{equation}
The task is therefore reduced to calculating $\kappa_\mathbf{k}^{\mu\alpha_1\hdots\alpha_n}$ via Feynman diagrams. We will do this for the example cases of the linear conductivity and SHG conductivity, and note the cancellations that occur to eliminate the apparent $1/\omega$ and $1/\omega^2$ divergences, respectively.

\subsubsection{Length gauge}

In length gauge, one simply has the perturbation,
\begin{equation}
    \hat{V}(t) = e\,\hat{r}^\alpha E_\alpha(t)
    \text{,}
\end{equation}
where $\hat{r}^\alpha$ denotes the position operator. Care needs to be taken when interpreting the $\mathbf{k}$-space form of the operator $\hat{r}^\alpha$ \cite{Blount1962}. When calculating commutators with other operators, $\hat{r}^\alpha$ acts as a \textit{covariant derivative},
\begin{equation}
    \left[\hat{r}^\alpha,\hat{\mathcal{O}}\right]_{ab} \equiv i\left[\hat{\mathcal{D}}^\alpha,\hat{\mathcal{O}}\right]_{ab} = i\partial^\alpha\hat{\mathcal{O}}_{ab} + \left[\hat{\mathcal{A}}^\alpha,\hat{\mathcal{O}}\right]_{ab}
    \text{,}
\end{equation}
where $\mathcal{A}^\alpha_{ab}(\mathbf{k}) \equiv \braket{u_a(\mathbf{k})\vert\partial^\alpha u_b(\mathbf{k})}$ is the non-Abelian Berry connection. With this definition, nonlinear conductivity tensors in length gauge can be computed directly as (suppressing frequency dependence),
\begin{equation}
    \sigma^{\mu\alpha_1\hdots\alpha_n} = -\frac{e^{n+1}}{\hslash^n}\int[d\mathbf{k}]\,\text{tr}\left\{\hat{v}^\mu\cdot\hat{\rho}^{(n)}_{\hat{r}^{\alpha_1}\hdots\hat{r}^{\alpha_n}}\right\}
    \text{,}
\end{equation}
with,
\begin{equation}
    \hat{\rho}^{(n)}_{\hat{r}^{\alpha_1}\hdots\hat{r}^{\alpha_n}} = \frac{1}{\omega^{(n)}-\hat{\epsilon}}\circ\left[\hat{r}^{\alpha_n},\hat{\rho}^{(n-1)}_{\hat{r}^{\alpha_1}\hdots\hat{r}^{\alpha_{n-1}}}\right]
    \text{,}
\end{equation}
without need for extra terms as in the diagrammatic scheme for velocity gauge. While the length gauge expression is analytically more compact, it is numerically less straightforward to use, since $\mathbf{k}$-derivatives act iteratively on the density matrix. In velocity gauge, all the $\mathbf{k}$-derivatives act on the Hamiltonian, so they can be computed ahead of time to speed up calculations.

\subsection{Examples}

\subsubsection{Linear conductivity}

\begin{figure}
    \centering
    \vspace{0.3in}
    \includegraphics[width=0.9\linewidth]{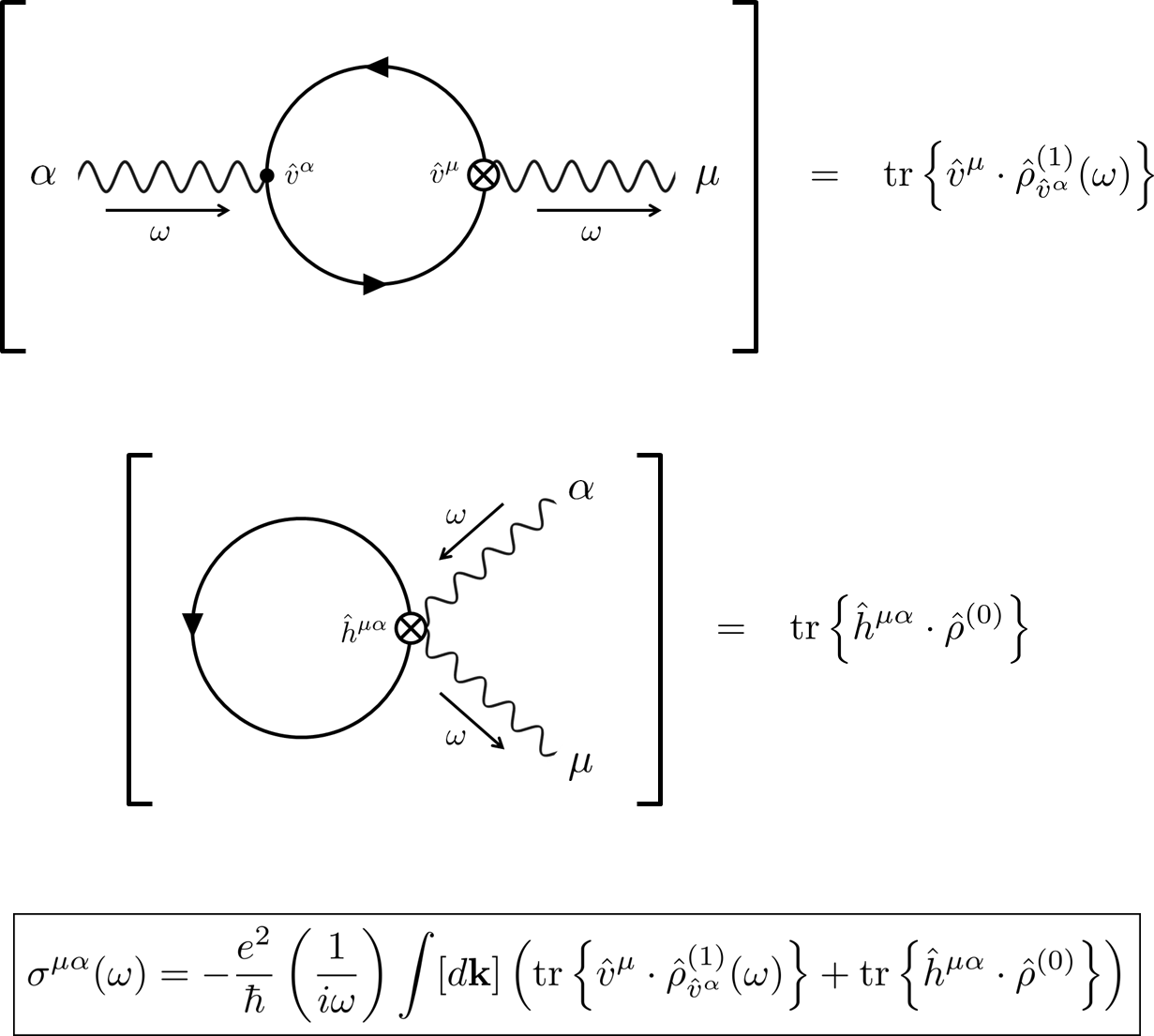}
    \vspace{0.15in}
    \caption{Diagrams relevant to computing the linear conductivity tensor $\sigma^{\mu\alpha}(\omega)$ in velocity gauge.}
    \label{fig:lin diagrams}
\end{figure}

The linear conductivity $\sigma^{\mu\alpha}(\omega;\omega)\equiv \sigma^{\mu\alpha}(\omega)$ is defined as follows,
\begin{equation}
    \left\langle \hat{j}^\mu(\omega) \right\rangle = \sigma^{\mu\alpha}(\omega) E_\alpha(\omega)
    \text{.}
\end{equation}
The calculation in velocity gauge can be written as the sum of two terms, represented diagrammatically in Figure \ref{fig:lin diagrams},
\begin{equation*}
    \sigma^{\mu\alpha}(\omega) = -\frac{e^2}{\hslash}\left(\frac{1}{i\omega}\right)\int[d\mathbf{k}]\,\kappa^{\mu\alpha}_\mathbf{k}(\omega)
    \text{,}
\end{equation*}
\begin{equation}
    \kappa_\mathbf{k}^{\mu\alpha}(\omega) = \text{tr}\left\{\hat{v}^\mu\cdot\hat{\rho}^{(1)}_{\hat{v}^\alpha}(\omega)\right\}+\text{tr}\left\{\hat{h}^{\mu\alpha}\cdot\hat{\rho}^{(0)}\right\}
    \text{,}
\end{equation}
where each operator within the traces is understood to be evaluated at $\mathbf{k}$. Note that $\hat{v}^\mu \equiv \hat{h}^\mu$, so we will use these interchangeably. Physically, these two terms correspond to the well-known paramagnetic current and diamagnetic current, respectively. The apparent $1/\omega$ divergence in the conductivity formula is canceled when these two terms are summed and integrated over $\textbf{k}$. It is interesting to note that this cancellation \textit{does not occur} in the case of a superconductor, where the low-frequency divergence of the conductivity is physical and results in the linear-in-$\mathbf{A}$ current in the London equation.

Assuming $\rho^{(0)}_{ab} = f_a\delta_{ab}$, we can re-express the linear conductivity as follows,
\begin{equation}
    \sigma^{\mu\alpha}(\omega) = \frac{ie^2}{\hslash\omega}\int[d\mathbf{k}]\left(\sum_{ab}f_{ab}\frac{v^\mu_{ab}v^\alpha_{ba}}{\omega-\epsilon_{ba}} + \sum_a f_a h^{\mu\alpha}_{aa}\right)
    \text{,}
\end{equation}
which matches velocity gauge results from the literature \cite{Passos2018}.

In length gauge, recalling that $[\hat{r}^\alpha,\hat{\mathcal{O}}]_{ab} = i\partial^\alpha\mathcal{O}_{ab} + [\hat{\mathcal{A}}^\alpha,\hat{\mathcal{O}}]_{ab}$, where $\hat{\mathcal{A}}^\alpha$ is the non-Abelian Berry connection, we write,
\begin{equation}
\begin{split}
    \sigma^{\mu\alpha}_\text{len.}(\omega) &= -\frac{e^2}{\hslash}\int[d\mathbf{k}]\, \text{tr}\left\{\hat{v}^\mu\cdot\hat{\rho}^{(1)}_{\hat{r}^\alpha}(\omega)\right\} \\
    &=-\frac{e^2}{\hslash}\int[d\mathbf{k}]\left(\sum_{ab}f_{ab}\frac{v^\mu_{ab}\mathcal{A}^\alpha_{ba}}{\omega-\epsilon_{ba}}+\sum_a \frac{iv^\mu_{aa}\partial^\alpha f_a}{\omega}\right)
    \text{.}
\end{split}
\end{equation}
Using $v^\alpha_{ab} = i\epsilon_{ab}\mathcal{A}^\alpha_{ab}$ for $a \neq b$, this allows us to write,
\begin{equation}
\begin{split}
    \sigma^{\mu\alpha}_\text{len.}(\omega) =\frac{ie^2}{\hslash}\int[d\mathbf{k}]&\left(\sum_{ab}f_{ab}\frac{v^\mu_{ab}v^\alpha_{ba}}{\epsilon_{ba}(\omega-\epsilon_{ba})}\right.\\
    &\left.-\sum_a \frac{v^\mu_{aa}\partial^\alpha f_a}{\omega}\right)
    \text{.}
\end{split}    \label{len lin 2}
\end{equation}

\subsubsection{Second harmonic generation conductivity}

\begin{figure*}[t]
    \centering
    \includegraphics[width=\textwidth]{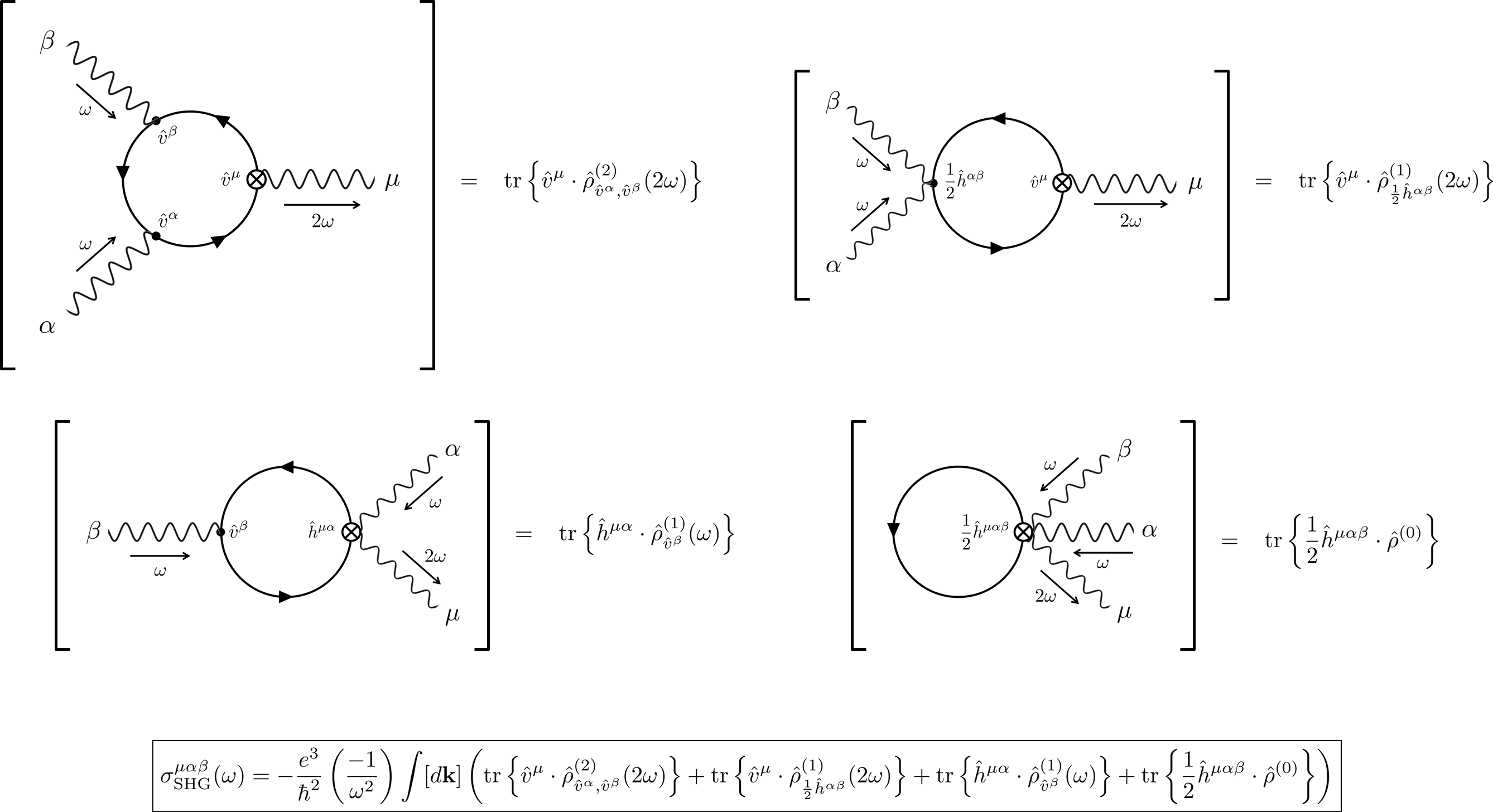}
    \vspace{0in}
    \caption{Diagrams relevant to computing the SHG conductivity tensor $\sigma_\text{SHG}^{\mu\alpha\beta}(\omega)$ in velocity gauge.}
    \label{fig:SHG diagrams}
\end{figure*}

The second harmonic generation conductivity $\sigma^{\mu\alpha\beta}(2\omega;\omega,\omega) \equiv \sigma^{\mu\alpha\beta}_\text{SHG}(\omega)$ is defined as follows,
\begin{equation}
    \left\langle\hat{j}^\mu(2\omega)\right\rangle = \sigma^{\mu\alpha\beta}_\text{SHG}(\omega)E_\alpha(\omega)E_\beta(\omega)
\end{equation}
The calculation in velocity gauge can be written as the sum of four terms, represented diagrammatically in Figure \ref{fig:SHG diagrams},
\begin{equation*}
    \sigma^{\mu\alpha\beta}_\text{SHG}(\omega) = -\frac{e^3}{\hslash^2}\left(\frac{-1}{\omega^2}\right)\int[d\mathbf{k}]\,\kappa^{\mu\alpha\beta}_{\text{SHG},\mathbf{k}}(\omega)
    \text{,}
\end{equation*}
\begin{equation}
\begin{split}
    \kappa^{\mu\alpha\beta}_{\text{SHG},\mathbf{k}}(\omega) =\,\,& \text{tr}\left\{\hat{v}^\mu\cdot\hat{\rho}^{(2)}_{\hat{v}^\alpha,\hat{v}^\beta}(2\omega)\right\} + \text{tr}\left\{\hat{v}^\mu\cdot\hat{\rho}^{(1)}_{\frac{1}{2}\hat{h}^{\mu\alpha}}(2\omega)\right\} \\
    &+\text{tr}\left\{\hat{h}^{\mu\alpha}\cdot\hat{\rho}^{(1)}_{\hat{v}^\beta}(\omega)\right\} + \text{tr}\left\{\frac{1}{2}\hat{h}^{\mu\alpha\beta}\cdot\hat{\rho}^{(0)}\right\}
    \text{.}
\end{split}
\end{equation}
Assuming $\rho^{(0)}_{ab} = f_a\delta_{ab}$, this becomes,
\begin{equation}
\begin{split}
    \kappa^{\mu\alpha\beta}_{\text{SHG},\mathbf{k}}(\omega) =\,\,& \sum_{a,b,c}v^\mu_{ab}\frac{\displaystyle v^\beta_{bc}\frac{v^\alpha_{ca}f_{ac}}{\omega-\epsilon_{ca}}-\frac{f_{cb}v^\alpha_{bc}}{\omega-\epsilon_{bc}}v^\beta_{ca}}{2\omega-\epsilon_{ba}} \\
    &+\sum_{a,b}v^\mu_{ab}\frac{\frac{1}{2}h^{\alpha\beta}_{ba}f_{ab}}{2\omega-\epsilon_{ab}} + \sum_{a,b}h^{\mu\alpha}_{ab}\frac{v^\beta_{ba}f_{ab}}{\omega-\epsilon_{ba}} \\
    &+\sum_{a}\frac{1}{2}h^{\mu\alpha\beta}_{aa} f_a
    \text{.}
\end{split}
\end{equation}
We note that, under $\mathbf{k}$-space symmetry considerations alone, the contributions from the first and last diagrams in Figure \ref{fig:SHG diagrams} should vanish under $\mathbf{k}$-space integration over the full Brillouin zone (that is, their integrands are odd in $\mathbf{k}$ while the integration region is symmetric in $\mathbf{k}$). Therefore, the content of the cancellation required to eliminate the $1/\omega^2$ prefactor in the conductivity comes solely from the second and third diagrams, just as the cancellation in the linear conductivity comes from only two diagrams.

\section{Calculations of spatially dispersive nonlinear optical responses}

\subsection{Generalities}

A nonlinear optical conductivity tensor is in full generality a function of the wavevectors of the input/output optical fields $\mathbf{q}_i$, but is typically treated in the $\mathbf{q}\to 0$ limit. This is justified within the so-called dipole approximation, in which it is assumed that the length scale of the electronic states being coupled by optical transitions (for instance, the size of electronic orbitals) is very small compared to the wavelength of the field driving the transition. However, there are many cases where this approximation should be expected to fail (for instance, in band systems whose Wannier representation has poor localization). For these reasons, one may be interested in the lowest-order contribution of the spatial variation of the optical fields,
\begin{equation}
    E_{\alpha_j}(\mathbf{r},t) = E_{\alpha_j}(\mathbf{0},t)\left(1-i\mathbf{q}_j\cdot\hat{\mathbf{r}}\right) + \mathcal{O}(\mathbf{q}^2)
    \text{.}
\end{equation}

Previous works \cite{Malashevich2010, Ahn2022} have identified how to expand the linear optical conductivity to first-order in $\mathbf{q}$. Interpreting the zeroth-order contribution to the current operator as the time-derivative of the electric dipole, $\hat{P}^\mu = -e \hat{r}^\mu$, the first-order terms include contributions from the electric quadrupole and magnetic dipole, \cite{Melrose2009}
\begin{equation}
    \hat{j}^\mu = \partial_t\hat{P}^\mu - iq_\nu\left(\frac{1}{2}\partial_t\hat{Q}^{\nu\mu} + c\epsilon^{\nu\mu\rho}\hat{M}_\rho\right) + \mathcal{O}(\mathbf{q}^2)
    \text{,} \label{j q}
\end{equation}
where $\hat{Q}^{\nu\mu} = -e \hat{r}^\nu\hat{r}^\mu$ is the electric quadrupole operator, and $\hat{M}_\rho = -\frac{e}{2c}\epsilon_{\rho\alpha\beta}\hat{r}^\alpha\hat{j}^\beta$ is the magnetic dipole operator. Eq. (\ref{j q}) constitutes the starting point for a velocity gauge form for light-matter coupling up to first order in $\mathbf{q}$. In this work, we will focus on the electric quadrupole contribution, and determine the velocity gauge and length gauge expressions for treating this contribution. In length gauge, the perturbation takes the form,
\begin{equation}
    \hat{V}(\omega) = \left(\hat{r}^\alpha-i q_\nu \left(\frac{1}{2}\hat{Q}^{\nu\alpha} + \frac{c}{\omega} \epsilon^{\nu\mu\rho}\hat{M}_\rho\right)\right) E_\alpha(\omega)
\end{equation}
We constrain the exact form of the electric quadrupole operator by ensuring that we recover $-\frac{\partial}{\partial r^\alpha}V(\omega) = E_\alpha(\omega)(1-iq_\nu r^\nu) + \mathcal{O}(\mathbf{q}^2)$. Notably, this gives a relative factor of 2 to the off-diagonal spatial components of $\hat{Q}^{\nu\mu}$. For example, in $d=2$, we have,
\begin{equation}
    \hat{Q}^{\nu\mu} =
    \begin{pmatrix}
        \hat{Q}^{xx} & \hat{Q}^{xy} \\
        \hat{Q}^{yx} & \hat{Q}^{yy}
    \end{pmatrix}
    =
    \begin{pmatrix}
        \hat{\mathcal{A}}^{x}\cdot\hat{\mathcal{A}}^{x} & 2\hat{\mathcal{A}}^{x}\cdot\hat{\mathcal{A}}^{y} \\
        2\hat{\mathcal{A}}^{y}\cdot\hat{\mathcal{A}}^{x} & \hat{\mathcal{A}}^{y}\cdot\hat{\mathcal{A}}^{y}
    \end{pmatrix}
    \text{,}
\end{equation}
where we recall the non-Abelian Berry connection $\mathcal{A}^{\mu}_{ab} \equiv \left\langle u_{a\mathbf{k}} \right. \left| \partial^\mu u_{b\mathbf{k}} \right\rangle$.

\begin{figure}[ht]
    \centering
    \includegraphics[width=\linewidth]{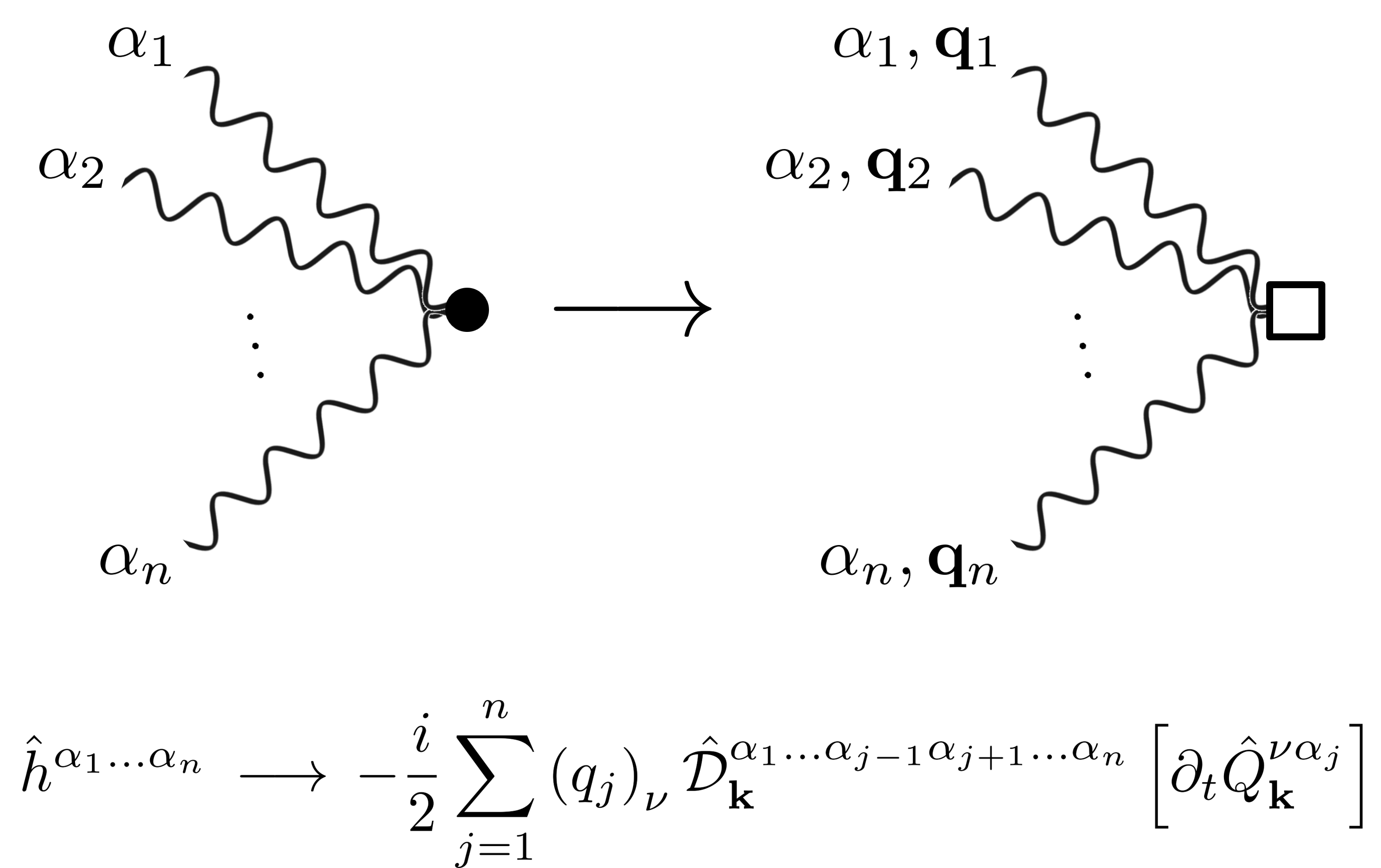}
    \caption{Rule for determining the first-order-in-$\mathbf{q}$ contribution from a vertex operator. We change a vertex from a dot to a square to indicate that it is being perturbed to first order in $\mathbf{q}$.}
    \label{fig:feynmanrulesq}
\end{figure}

It has been shown \cite{Malashevich2010} that the first-order-in-$\mathbf{q}$ contribution to the current operator matrix element can be cast in terms of the following object,
\begin{equation}
\begin{split}
    \left\langle u_{n,\mathbf{k}+\frac{\mathbf{q}}{2}}\right\vert \hat{v}^\mu_\mathbf{k} \left\vert u_{m,\mathbf{k}-\frac{\mathbf{q}}{2}}\right\rangle &= -\frac{i}{2}q_\nu\left\{\hat{A}^\nu_\mathbf{k},\hat{v}^{\mu}_\mathbf{k}\right\}_{nm} \\
    &\equiv -\frac{i}{2}q_\nu\, \left(\partial_t \hat{Q}^{\nu\mu}_\mathbf{k}\right)_{nm}
\end{split}
\end{equation}
where in the last line we make the crucial identification that this should equal the time derivative of the electric quadrupole moment. This is the basis for a new vertex rule we introduce, illustrated in Figure \ref{fig:feynmanrulesq} and summarized in the following expression for the velocity-gauge perturbation,
\begin{equation}
\begin{split}
    \hat{V}_\mathbf{k}(t) = &\sum_{n=1}^\infty \frac{e^n}{n!} A_{\alpha_1}(t) \hdots A_{\alpha_n}(t) \\
    &\times \Bigg(\hat{\mathcal{D}}^{\alpha_1\hdots\alpha_n}_\mathbf{k}\left[\hat{\mathcal{H}}_\mathbf{k}\right] \\
    &+\frac{1}{2}\sum_{j=1}^n(q_j)_\nu\hat{\mathcal{D}}_\mathbf{k}^{\alpha_1\hdots\alpha_{j-1}\alpha_{j+1}\hdots\alpha_n}\left[\hat{\mathcal{H}}_\mathbf{k} \right.,\left. \hat{Q}^{\nu\alpha_j}_\mathbf{k}\right]\Bigg)    
    \text{,}\label{new V}
\end{split}
\end{equation}
where the vectors $\mathbf{q}_j$ are the wavevectors from the $n$ fields $A_{\alpha_1}(t),\hdots, A_{\alpha_n}(t)$. We can interpret this as introducing an additional type of vertex in the diagrammatic method, one that takes into account at first order the momentum ($\hslash\mathbf{q}_j$) of each photon connected to that vertex. We will use the terms ``momentum" and ``wavevector" interchangeably.

Schematically, this leads to a spatially dispersive correction to nonlinear optical conductivities,
\begin{equation}
    \sigma^{\mu\alpha_1\hdots\alpha_n}(\mathbf{q}) = \sigma^{\mu\alpha_1\hdots\alpha_n}_{(0)} + q_\nu\sigma^{\nu\mu\alpha_1\hdots\alpha_n}_{(1)} + \mathcal{O}(\mathbf{q}^2)
    \text{,}
\end{equation}
where $\mathbf{q}$ is some characteristic momentum of the input fields. To declutter notation, we have suppressed the $\omega$ dependence, which is understood to be $(\omega_{(n)};\omega_1\hdots\omega_n)$ in general. As before, in velocity gauge, we write this as,
\begin{equation}
    \sigma^{\nu\mu\alpha_1\hdots\alpha_n}_{(1)} = -\frac{e^{n+1}}{\hslash^n}\prod_{j=1}^n\left(\frac{1}{i\omega_j}\right) \int[d\mathbf{k}]\,\kappa_{(1),\mathbf{k}}^{\nu\mu\alpha_1\hdots\alpha_n}
    \text{,}
\end{equation}
and calculate $\kappa^{\nu\mu\alpha_1\hdots\alpha_n}_{(1),\mathbf{k}}$ using a diagrammatic scheme. As a consequence of Eq. (\ref{new V}),  this diagrammatic scheme amounts to taking each diagram for the $\mathcal{O}(\mathbf{q}^0)$ response and, for each vertex $\hat{h}^{\alpha_1\hdots\alpha_n}$, creating a new diagram with the following replacement rule (dropping $\mathbf{k}$ dependence for brevity),
\begin{equation}
\begin{split}
    \hat{h}^{\alpha_1\hdots\alpha_n} &\to \frac{1}{2}\sum_{j=1}^n (q_j)_\nu\hat{\mathcal{D}}^{\alpha_1\hdots\alpha_{j-1}\alpha_{j+1}\hdots\alpha_n}\left[\hat{\mathcal{H}}\right.,\left.\hat{Q}^{\nu\alpha_j}\right] \\
    &\equiv -\frac{i}{2}\sum_{j=1}^n(q_j)_\nu \partial_t\hat{Q}^{\nu\alpha_j;\alpha_1\hdots\alpha_{j-1}\alpha_{j+1}\hdots\alpha_n}
    \text{,} \label{replacement rule}
\end{split}
\end{equation}
where $(q_j)_\nu$ is the $\nu$ component of the wavevector for the photon with polarization $\alpha_j$ (again, summation over the repeated index $\nu$ is implied). In the second line, we introduce a compact notation, whereby indices after the semicolon indicate covariant $\mathbf{k}$-derivatives. To make sense of Eq. (\ref{replacement rule}), note that for a single-photon vertex, the replacement rule reduces to,
\begin{equation}
    \hat{h}^{\alpha_1} \to -\frac{i}{2}(q_1)_\nu \partial_t\hat{Q}^{\nu\alpha_1}
    \text{.}
\end{equation}
In other words, it replaces an \textit{electric dipole} coupling $\hat{h}^{\alpha_1} \equiv \hat{v}^{\alpha_1} \equiv \partial_t\hat{r}^{\alpha_1}$ with an \textit{electric quadrupole} coupling $\partial_t\hat{Q}^{\nu\alpha_1}$. The extension to multi-photon vertices can then be understood as a sum of terms that change the coupling from dipolar to quadrupolar for \textit{one photon at a time} while having the rest of the photons contribute covariant $\mathbf{k}$-derivatives of the resulting quadrupolar current operator. We illustrate this new vertex rule in Figure \ref{fig:feynmanrulesq}.

We note that the output vertex should also be perturbed to linear-in-$\mathbf{q}$ order in this scheme, and we denote this by a ``$\boxtimes$" symbol in the diagrams. As in the spatially uniform case, the only special treatment that must be given to the output vertex is that the symmetry factor $1/n!$ should only include the $n$ input photons (i.e. excluding the output photon, for whom we reserve the index $\mu$ to denote its polarization).

\subsection{Examples}

\subsubsection{Spatially dispersive linear conductivity}

\begin{figure}[h]
    \centering
    \vspace{0.3in}
    \includegraphics[width=0.65\linewidth]{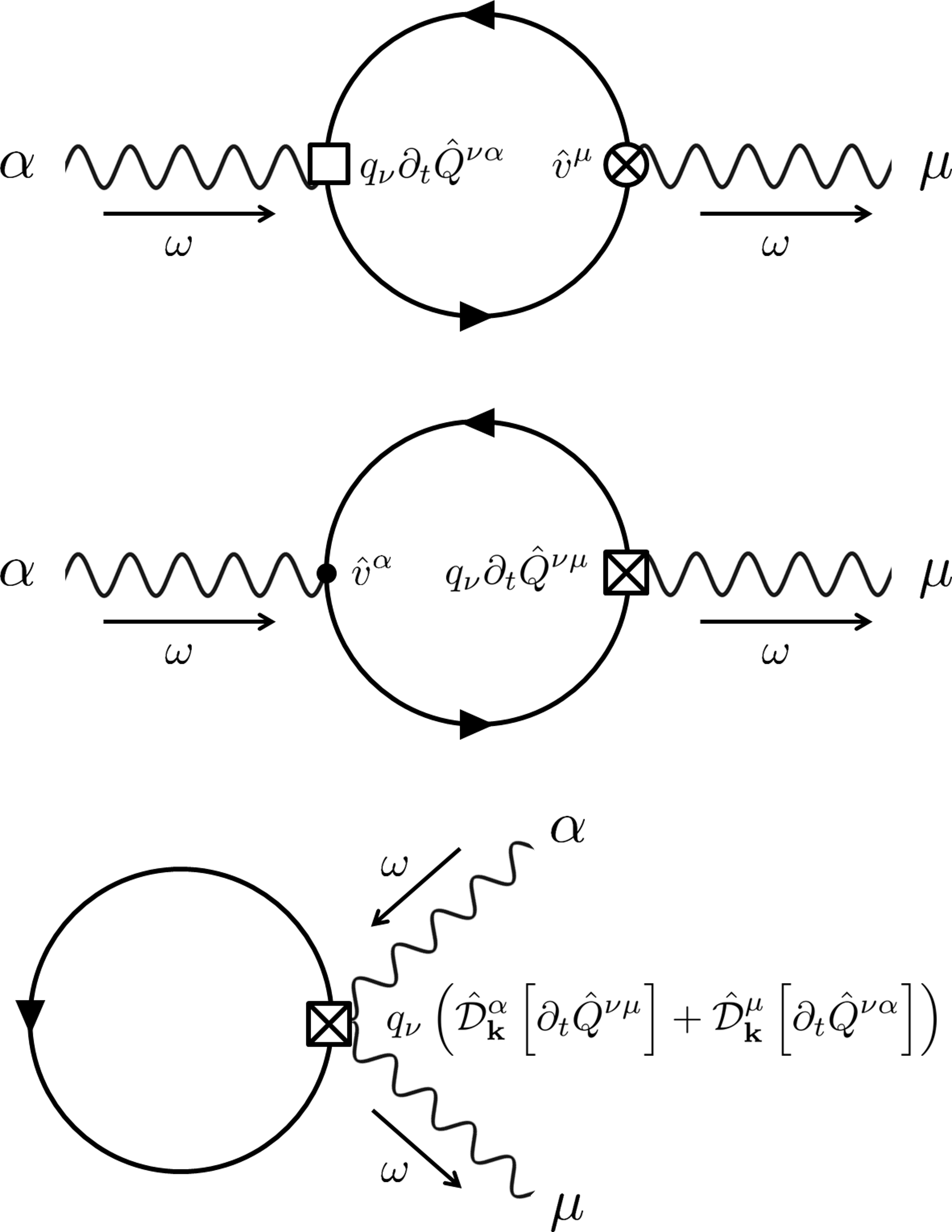}
    \vspace{0.15in}
    \caption{Diagrams relevant to computing the \textit{spatially dispersive} linear conductivity tensor to lowest order ($\sigma_{(1)}^{\nu\mu\alpha}(\omega)$) in velocity gauge. The third diagram is shown for completeness, but has a vanishing contribution since it only contains a vertex with net-zero momentum flow. (We omit a common factor of $-\frac{i}{2}$ from each diagram.)}
    \label{fig:qlin diagrams}
\end{figure}

Applying the Feynman rules in Figure \ref{fig:feynmanrulesq} to the linear conductivity diagrams in Figure \ref{fig:lin diagrams}, we obtain the diagrams in Figure \ref{fig:qlin diagrams}. These diagrams represent the following two terms,

\begin{equation}
\begin{split}
    \kappa_{(1),\mathbf{k}}^{\nu\mu\alpha}(\omega) = -\frac{i}{2}&\Big( \text{tr}\left\{\hat{v}^\mu\cdot\hat{\rho}^{(1)}_{\partial_t\hat{Q}^{\nu\alpha}}(\omega)\right\} \\
    &+ \text{tr}\left\{\partial_t\hat{Q}^{\nu\mu}\cdot\hat{\rho}^{(1)}_{\hat{v}^\alpha}(\omega)\right\} \\
    &+ \text{tr}\left\{\left(\partial_t\hat{Q}^{\nu\mu;\alpha}+\partial_t\hat{Q}^{\nu\alpha;\mu}\right)\cdot\hat{\rho}^{(0)}\right\}\Big)
    \text{.}
\end{split}
\end{equation}
The conductivity can therefore be written as,
\begin{equation}
\begin{split}
    \sigma^{\nu\mu\alpha}_{(1)}(\omega) = \frac{e^2}{2\hslash\omega} \int [d\mathbf{k}] &\Bigg\{\sum_{ab} f_{ab}\frac{v^\mu_{ab}\dot{Q}^{\nu\alpha}_{ba}+\dot{Q}^{\nu\mu}_{ab}v^\alpha_{ba}}{\omega-\epsilon_{ba}} \\
    &+\sum_a f_a \left(\dot{Q}^{\nu\mu;\alpha}_{aa} + \dot{Q}^{\nu\alpha;\mu}_{aa}\right)\Bigg\}
    \text{,}\label{qlin vel}
\end{split}
\end{equation}
where we abbreviate $\left\langle u_{a\mathbf{k}} \right|\partial_t \hat{Q}^{\nu\mu} \left|u_{b\mathbf{k}}\right\rangle \equiv \dot{Q}^{\nu\mu}_{ab} \equiv i\left[\hat{\mathcal{H}}\right.,\left.\hat{Q}^{\nu\mu}\right]_{ab}$. Despite the apparent divergence from the prefactor $1/\omega$, this response is regularized in the low-frequency limit, as we show numerically for an example model in Section \ref{Results}. To further justify this, we can re-derive this result in length gauge,
\begin{equation}
\begin{split}
    \sigma^{\nu\mu\alpha}_{(1)}(\omega) = \frac{ie^2}{2\hslash}\int[d\mathbf{k}]&\left(\text{tr}\left\{\hat{v}^\mu\cdot\hat{\rho}^{(1)}_{\hat{Q}^{\nu\alpha}}(\omega)\right\}\right. \\
    &\left.+\text{tr}\left\{\partial_t\hat{Q}^{\nu\mu}\cdot\hat{\rho}^{(1)}_{\hat{r}^\alpha}(\omega)\right\}\right)
    \text{.}\label{qlin len}
\end{split}
\end{equation}
We show numerically in Section \ref{Results} that Eqs. (\ref{qlin vel}) and (\ref{qlin len}) are equivalent.

\subsubsection{Spatially dispersive second harmonic generation conductivity}

\begin{figure*}[t]
    \centering
    \includegraphics[width=0.9\textwidth]{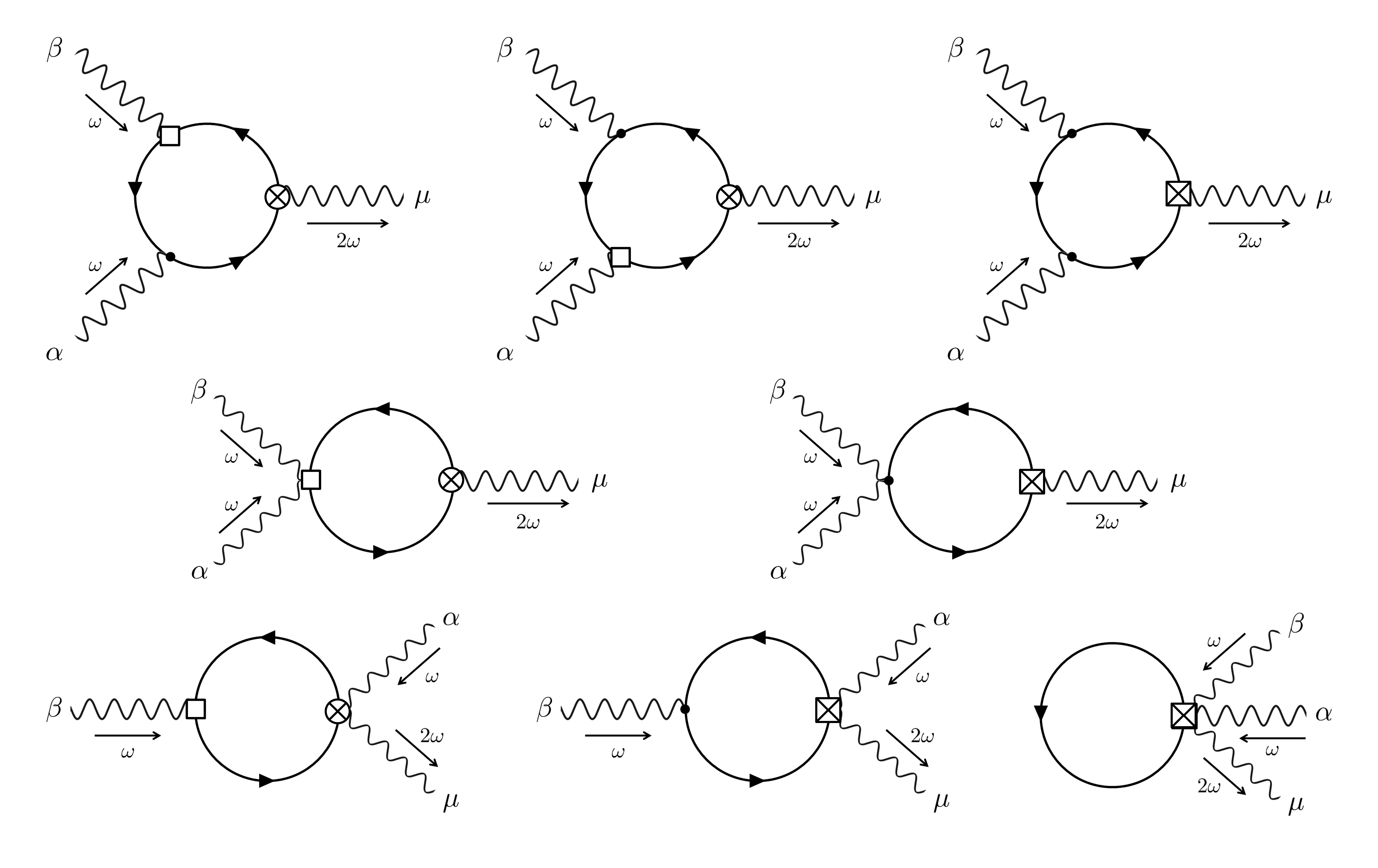}
    \caption{Diagrams for computing lowest-order \textit{spatially dispersive corrections} to the SHG conductivity tensor ($\sigma^{\nu\mu\alpha\beta}_{\text{SHG},(1)}(\omega)$) in velocity gauge.}
    \label{fig:qSHG diagrams}
\end{figure*}

Figure \ref{fig:qSHG diagrams} displays the eight diagrams for calculating the linear-in-$\mathbf{q}$ SHG conductivity. These yield the following seven terms,

\begin{equation}
\begin{split}
    \kappa^{\nu\mu\alpha\beta}_{\text{SHG},(1),\mathbf{k}}(\omega) &=\,-\frac{i}{2}\Bigg[ \text{tr}\left\{\hat{v}^\mu\cdot\hat{\rho}^{(2)}_{\hat{v}^\alpha,\partial_t\hat{Q}^{\nu\beta}}(2\omega)\right\} \\
    &+ \text{tr}\left\{\hat{v}^\mu\cdot\hat{\rho}^{(2)}_{\partial_t\hat{Q}^{\nu\alpha},\hat{v}^\beta}(2\omega)\right\} \\
    &+\text{tr}\left\{\partial_t\hat{Q}^{\nu\mu}\cdot\hat{\rho}^{(2)}_{\hat{v}^\alpha,\hat{v}^\beta}(2\omega)\right\} \\
    &+\text{tr}\left\{\hat{v}^\mu\cdot\hat{\rho}^{(1)}_{\frac{1}{2}\partial_t\hat{Q}^{\nu\alpha;\beta}+\frac{1}{2}\partial_t\hat{Q}^{\nu\beta;\alpha}}(2\omega)\right\} \\
    &+ \text{tr}\left\{\partial_t\hat{Q}^{\nu\mu}\cdot\hat{\rho}^{(1)}_{\frac{1}{2}\hat{h}^{\alpha\beta}}(2\omega)\right\} \\
    &+\text{tr}\left\{\hat{h}^{\mu\alpha}\cdot\hat{\rho}^{(1)}_{\partial_t\hat{Q}^{\nu\beta}}(\omega)\right\} \\
    &+ \text{tr}\left\{\left(\partial_t\hat{Q}^{\nu\mu;\alpha}+\partial_t\hat{Q}^{\nu\alpha;\mu}\right)\cdot\hat{\rho}^{(1)}_{\hat{v}^\beta}(\omega)\right\} \\
    &+ \text{tr}\left\{\frac{1}{2}\left(\partial_t\hat{Q}^{\nu\mu;\alpha\beta} + \text{[perms.]}\right)\cdot\hat{\rho}^{(0)}\right\}\Bigg]
    \text{,}
\end{split}
\end{equation}
where ``[perms.]" indicates the two other permutations of $\mu,\alpha,\beta$ modulo exchanges of the indices after the semicolon. Assuming $\rho^{(0)}_{ab} = f_a\delta_{ab}$, this becomes,
\begin{equation}
\begin{split}
    \kappa^{\nu\mu\alpha\beta}_{\text{SHG},(1),\mathbf{k}}(\omega) &=-\frac{i}{2}\sum_{a,b}\Bigg( v^\mu_{ab}\frac{N_{1,ba}^{\nu\alpha\beta}(\omega)}{2\omega-\epsilon_{ba}}-\dot{Q}^{\nu\mu}_{ab}\frac{N_{2,ba}^{\alpha\beta}(\omega)}{2\omega-\epsilon_{ba}} \\
    &\,\,\,\,\,\,\,\,\,\,\,\,\,\,\,\,\,\,\,\,\,\,\,\,\,\,\,\,\,+\frac{1}{2}v^\mu_{ab}\frac{(\dot{Q}^{\nu\alpha;\beta}_{ba}+\dot{Q}^{\nu\beta;\alpha}_{ba})f_{ab}}{2\omega-\epsilon_{ab}} \\
    &\,\,\,\,\,\,\,\,\,\,\,\,\,\,\,\,\,\,\,\,\,\,\,\,\,\,\,\,\,+\frac{1}{2}\dot{Q}^{\nu\mu}_{ab}\frac{h^{\alpha\beta}_{ba}f_{ab}}{2\omega-\epsilon_{ab}} +h^{\mu\alpha}_{ab}\frac{\dot{Q}^{\nu\beta}_{ba}f_{ab}}{\omega-\epsilon_{ba}} \\
    &\,\,\,\,\,\,\,\,\,\,\,\,\,\,\,\,\,\,\,\,\,\,\,\,\,\,\,\,\,+\left(\dot{Q}^{\nu\mu;\alpha}_{ab}+\dot{Q}^{\nu\alpha;\mu}_{ab}\right)\frac{v^\beta_{ba}f_{ab}}{\omega-\epsilon_{ba}}\Bigg) \\
    &\,\,\,\,\,-\frac{i}{4}\sum_{a}\left(\dot{Q}^{\nu\mu;\alpha\beta}_{aa}+\dot{Q}^{\nu\alpha;\beta\mu}_{aa}+\dot{Q}^{\nu\beta;\mu\alpha}_{aa}\right)f_a
    \text{,}
\end{split}
\end{equation}
where,
\begin{equation}
\begin{split}
    N_{1,ba}^{\nu\alpha\beta}(\omega) &= \sum_c\left( \dot{Q}^{\nu\beta}_{bc}\frac{v^\alpha_{ca}f_{ac}}{\omega-\epsilon_{ca}}-\frac{f_{cb}v^\alpha_{bc}}{\omega-\epsilon_{bc}}\dot{Q}^{\nu\beta}_{ca}\right. \\
    &\,\,\,\,\,\,\,\,\,\,\,\,\,\,\,\,\,\,\,\,\,+\left.
    v^\beta_{bc}\frac{\dot{Q}^{\nu\alpha}_{ca}f_{ac}}{\omega-\epsilon_{ca}}-\frac{f_{cb}\dot{Q}^{\nu\alpha}_{bc}}{\omega-\epsilon_{bc}}v^\beta_{ca}\right)\text{,} \\
    N_{2,ba}^{\alpha\beta}(\omega) &= \sum_c\left(v^\beta_{bc}\frac{v^\alpha_{ca}f_{ac}}{\omega-\epsilon_{ca}}-\frac{f_{cb}v^\alpha_{bc}}{\omega-\epsilon_{bc}}v^\beta_{ca}\right)
    \text{.}
\end{split}
\end{equation}
In our forthcoming results, we compare to the following compact length gauge formula,
\begin{equation}
\begin{split}
    \sigma^{\nu\mu\alpha\beta}_{(1)}(\omega) = \frac{ie^3}{2\hslash^2}\int[d\mathbf{k}]&\Bigg(\text{tr}\left\{\hat{v}^{\mu}\cdot\hat{\rho}^{(2)}_{\hat{Q}^{\nu\alpha},\hat{r}^\beta}\right\} \\
    &+\text{tr}\left\{\hat{v}^{\mu}\cdot\hat{\rho}^{(2)}_{\hat{r}^\alpha,\hat{Q}^{\nu\beta}}\right\} \\
    &+\text{tr}\left\{\partial_t\hat{Q}^{\nu\mu}\cdot\hat{\rho}^{(2)}_{\hat{r}^\alpha,\hat{r}^\beta}\right\}\Bigg)
\text{.}
\end{split}
\end{equation}

\section{Results for minimal models}
\label{Results}

\begin{figure*}[t]
    \centering
    \includegraphics[width=\textwidth]{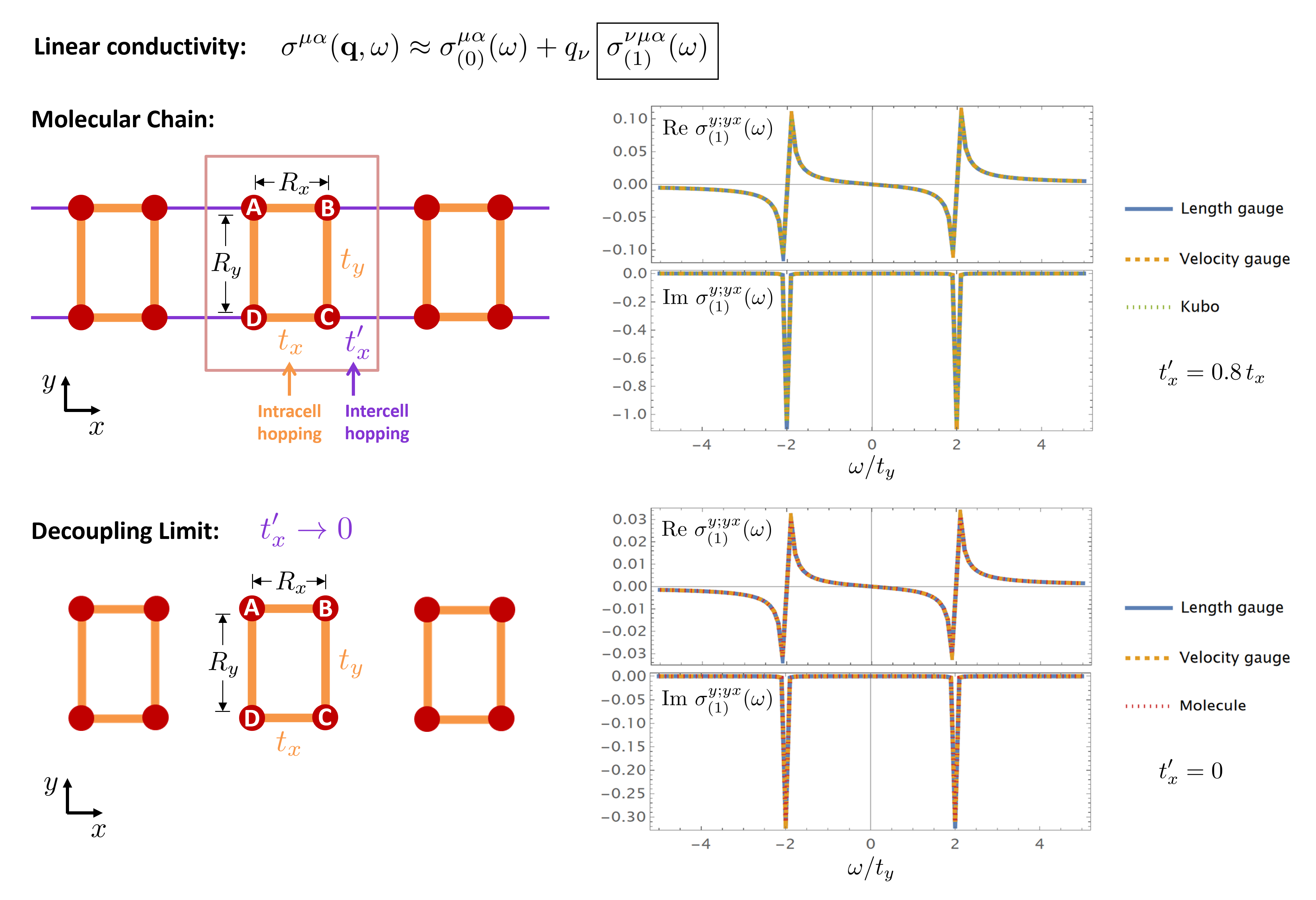}
    \caption{Results for the spatially dispersive correction to the \textit{linear conductivity} ($\sigma^{\nu\mu\alpha}_{(1)}(\omega)$) calculated on a minimal quasi-2D model. This model takes the form of a chain of rectangular molecules with inter-molecule couplings, with a sensible ``decoupling limit" in which the problem reduces to the well-understood problem of a multipole response in a molecule. (Top) Agreement is demonstrated for the molecular chain between our scheme (in both length and velocity gauge) and the expression determined in Ref. \cite{Malashevich2010} from the Kubo formula expanded to first order in $\mathbf{q}$. (Bottom) Agreement is also demonstrated in the decoupling limit between our scheme (in both length and velocity gauge) and the electric quadrupole linear response from an isolated molecule. Note that inversion-breaking is needed to observe this response, so we introduce a small static on-site potential ($\Delta$) of $\pm 0.1$ on sublattices $AD$/$BC$ to produce these results.}
    \label{fig:qLIN}
\end{figure*}

\begin{figure*}[t]
    \centering
    \includegraphics[width=\textwidth]{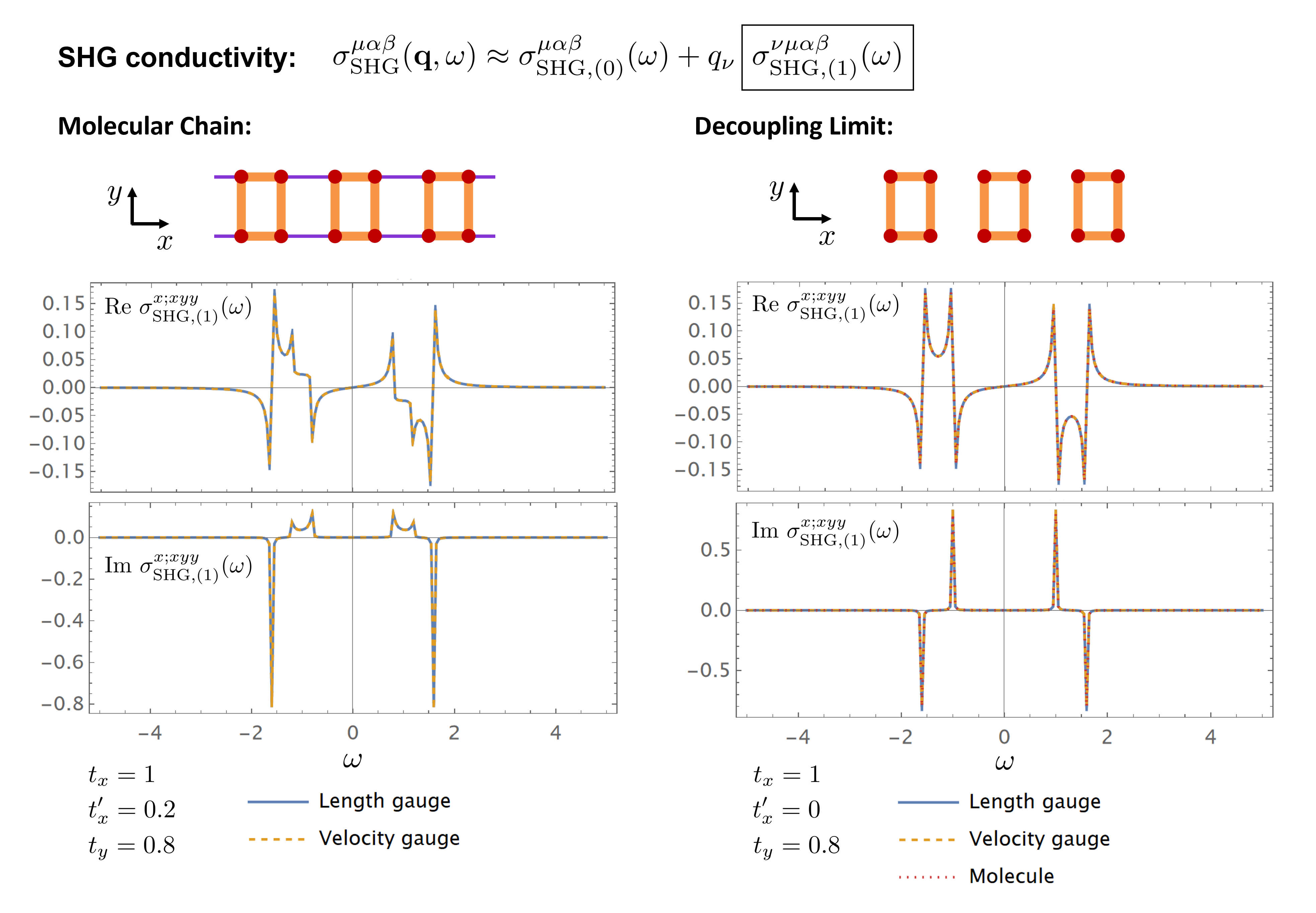}
    \caption{Results for the spatially dispersive correction to the \textit{second harmonic generation conductivity} ($\sigma^{\nu\mu\alpha\beta}_{\text{SHG},(1)}(\omega)$) calculated on the molecular chain model detailed in Figure \ref{fig:qLIN}. Agreement is shown between length gauge and velocity gauge in both the molecular chain and in the decoupling limit. Agreement is also shown with the analogous single-molecule calculation. We use $t_x'=0.8$ }
    \label{fig:qSHG}
\end{figure*}

In this section, we test our scheme on a minimal tight-binding model with three useful features: (1) inversion symmetry, (2) a quasi-2D structure allowing for multiple different polarizations, and (3) a clear ``decoupling limit" in which the model reduces to the well-understood problem of multipole optical responses in a molecule \cite{Cohen-Tannoudji1997, Melrose2009, Jaszunski2017}. We will confirm that, up to electric quadrupole order, a nonzero SHG response indeed exists in this centrosymmetric system, and we can therefore treat $\mathbf{q}$ on the same footing as a static inversion-breaking parameter in this model.

\subsection{Rectangular molecule}

We first establish the response up to first order in $\mathbf{q}$ and up to second order in the field strength for a 4-site rectangular molecule with only nearest-neighbor hoppings between the sites. Here, the position operators are unambiguous up to a choice of origin, and so one can straightforwardly describe the problem in length gauge. We write the Hamiltonian as,
\begin{equation}
    \hat{\mathcal{H}} = -
    \begin{pmatrix}
    0 & t_x & 0 & t_y \\
    t_x & 0 & t_y & 0 \\
    0 & t_y & 0 & t_x \\
    t_y & 0 & t_x & 0
    \end{pmatrix}
    \text{.} \label{H mol}
\end{equation}
Assume $t_x$ and $t_y$ are both positive, and without loss of generality, assume $t_x>t_y$. We use a basis in which the position operators $\hat{r}^{x}$ and $\hat{r}^{y}$ take the form,
\begin{equation}
    \hat{r}^{x} = R_x
    \begin{pmatrix}
    -1 & 0 & 0 & 0 \\
    0 & 1 & 0 & 0 \\
    0 & 0 & 1 & 0 \\
    0 & 0 & 0 & -1
    \end{pmatrix}
    ,\,\,\,
    \hat{r}^{y} = R_y
    \begin{pmatrix}
    1 & 0 & 0 & 0 \\
    0 & 1 & 0 & 0 \\
    0 & 0 & -1 & 0 \\
    0 & 0 & 0 & -1
    \end{pmatrix}
    \text{,} \label{r operators}
\end{equation}
where $R_x$ and $R_y$ denote half the length of the molecule in the $x$ and $y$ directions, respectively. Note that the quadrupole operators $\hat{r}^{x}\hat{r}^{x}$ and $\hat{r}^{y}\hat{r}^{y}$ are proportional to the identity matrix, so they cannot mediate any nonzero responses in the density matrix equation of motion. On the other hand, $\hat{r}^{x}\hat{r}^{y}\equiv\hat{r}^{y}\hat{r}^{x} = R_x R_y \text{diag}(-1, 1, -1, 1)$, which \textit{can} mediate a nonzero response. So one should only expect a response when the electric field is perpendicular to the incident $\mathbf{q}$ vector, which is sensible.

This Hamiltonian is diagonalized by a unitary transformation such as,
\begin{equation}
    \hat{\mathcal{U}} = \frac{1}{2}
    \begin{pmatrix}
    1 & 1 & 1 & 1 \\
    1 & 1 & -1 & -1 \\
    1 & -1 & -1 & 1 \\
    1 & -1 & 1 & -1
    \end{pmatrix}
    \text{,}
\end{equation}
yielding levels with energies $-t_x-t_y$, $-t_x+t_y$, $t_x-t_y$, $t_x+t_y$, in ascending order. We calculate the electric quadrupole SHG response using,
\begin{equation}
\begin{split}
    \sigma^{\nu\mu\alpha\beta}_{\text{SHG},(1)}(\omega) &= \frac{ie^3}{2\hslash^2}\bigg(\text{tr}\left\{\hat{v}^\mu\cdot\hat{\rho}^{(2)}_{\hat{Q}^{\nu\alpha},\hat{r}^\beta}(2\omega)\right\} \\
    &\,\,\,\,\,\,\,\,\,\,\,\,\,\,\,\,\,\,\,\,\,\,+\text{tr}\left\{\hat{v}^\mu\cdot\hat{\rho}^{(2)}_{\hat{r}^\alpha,\hat{Q}^{\nu\beta}}(2\omega)\right\} \\
    &\,\,\,\,\,\,\,\,\,\,\,\,\,\,\,\,\,\,\,\,\,\,+\text{tr}\left\{\partial_t\hat{Q}^{\nu\mu}\cdot\hat{\rho}^{(2)}_{\hat{r}^\alpha,\hat{r}^{\beta}}(\omega)\right\}\bigg)
    \text{,}
\end{split}
\end{equation}
starting from an initial density matrix of $\hat{\rho}^{(0)} = \text{diag}(f_1,f_2,f_3,f_4)$. We find,
\begin{equation}
    \sigma^{xxyy}_{\text{SHG},(1)}(\omega) = -\frac{4ie^3}{\hslash^2}\frac{(f_{12}-f_{34}) t_x t_y \omega R_x^2 R_y^2}{(\omega^2-t_x^2)(\omega^2-(2t_y)^2)}
    \text{.}\label{qSHG molecule 1}
\end{equation}
\begin{equation}
    \sigma^{yyxx}_{\text{SHG},(1)}(\omega) = -\frac{4ie^3}{\hslash^2}\frac{(f_{12}-f_{34}) t_x t_y \omega R_x^2 R_y^2}{(\omega^2-(2t_x)^2)(\omega^2-t_y^2)}
    \text{.}\label{qSHG molecule 2}
\end{equation}
This result reveals a number of interesting features. First of all, it is an example of a nonzero second harmonic response in an inversion-symmetric system, where the inversion breaking is allowed by the nonzero order in $\mathbf{q}$. Secondly, the response intriguingly vanishes for half-filling, owing to a symmetry between the Bloch states in the occupied and unoccupied energy levels. We note that introducing diagonal hoppings breaks this symmetry, and therefore allows a nonzero response at half-filling, at the expense of the energetic factors in Eq. (\ref{qSHG molecule 1})-(\ref{qSHG molecule 2}) becoming considerably more complicated. In other words, the response takes the form,
\begin{equation}
    \sigma^{xxyy}_{\text{SHG},(1)}(\omega) = -\frac{ie^3}{\hslash^2} F(\omega) R_x^2 R_y^2
    \text{,}
\end{equation}
where $F(\omega)$ only depends on $\omega$, the hopping amplitudes, and the Fermi factors.

Appealingly, $\sigma^{\nu\mu\alpha\beta}_{\text{SHG},(1)}(\omega)$ can be seen to depend on four factors of the length of a molecule, in the directions specified by $\mathbf{q}$, $\mathbf{j}$, and the two $\mathbf{E}$-fields. Therefore, a larger molecule exhibits a larger magnitude for this spatially dispersive correction. The size of the molecule can be considered as a proxy for the coherence length of electronic states. Crucially, in the case of obstructed band systems forcing delocalized Wannier states, this characteristic coherence length may be on the order of many unit cells, and therefore one should expect substantial spatially dispersive corrections from these systems.

\subsection{Chain of rectangular molecules}

Transitioning to a band theory, we study a quasi-1D model in which two copies of the Rice-Mele model form the legs of a ladder. We define the Hamiltonian as,
\begin{equation}
    \hat{\mathcal{H}}^\text{o}_k = -
    \begin{pmatrix}
        \Delta & g_\text{o}(k) & 0 & t_y \\
        g_\text{o}^*(k) & -\Delta & t_y & 0 \\
        0 & t_y & -\Delta & g_\text{o}^*(k) \\
        t_y & 0 & g_\text{o}(k) & \Delta
    \end{pmatrix}
    \text{,}
\end{equation}
\begin{equation}
    g_\text{o}(k) = t_x + t_x' e^{-ik}
    \text{.}
\end{equation}
This Hamiltonian reduces to the molecule Hamiltonian in Eq. (\ref{H mol}) in the limit $t_x'\to 0$ (and $\Delta\to 0$). The purpose of the parameter $\Delta$ (which breaks inversion if $t_x\neq t_x'$) is to allow calculation of the spatially dispersive correction to the \textit{linear} conductivity (which vanishes under inversion symmetry). Here, we assume translational symmetry in the $x$-direction, with crystal momentum $k \in [0,2\pi)$. 

We have written the Hamiltonian in what we will call ``orbital form" (hence the subscript ``o"), in which it has the desirable property of being periodic with respect to $k$-space translations by a reciprocal lattice vector,
\begin{equation}
    \hat{\mathcal{H}}^\text{o}_k = \hat{\mathcal{H}}^\text{o}_{k+2\pi}
    \text{.}
\end{equation}
However, this comes at the cost of neglecting Bloch phases $e^{i k \Delta x}$ for \textit{intracell} hoppings, and hence the normal Peierls substitution $k \to k + eA(t)$ with this Hamiltonian will neglect intracell currents. We will call this form of the Hamiltonian ``orbital form," (hence the superscript ``o") since it may be viewed as treating all states within the unit cell as orbitals on a single generalized site in the unit cell. There are two equivalent options for dealing with this. One is to define the position operator with an intercell and intracell part,
\begin{equation}
    \hat{r}_i \to \hat{r}_i + \hat{\tau}
    \text{,}
\end{equation}
and to explicitly add the intracell part of the velocity operator $\hat{v}_\text{intra} \equiv i\left[\hat{\mathcal{H}}, \hat{\tau}\right]$ when calculating currents. Alternatively, one can define a form of the $k$-space Hamiltonian such that the Bloch phases \textit{do} accurately reflect intracell processes,
\begin{equation}
    \hat{\mathcal{H}}^\text{s}_k = -
    \begin{pmatrix}
        \Delta & g_\text{s}(k) & 0 & t_y \\
        g_\text{s}^*(k) & -\Delta & t_y & 0 \\
        0 & t_y & -\Delta & g_\text{s}^*(k) \\
        t_y & 0 & g_\text{s}(k) & \Delta
    \end{pmatrix}\label{sublattice form}
    \text{,}
\end{equation}
\begin{equation}
    g_\text{s}(k) = t_xe^{ik/2} + t_x' e^{-ik/2}
    \text{,}
\end{equation}
and find the unitary transformation $\hat{\mathcal{U}}^{\text{s}\to\text{o}}_k$ that takes this Hamiltonian back to the orbital form,
\begin{equation}
    \hat{\mathcal{H}}^\text{o}_k = \hat{\mathcal{U}}^{\text{s}\to\text{o}\dagger}_k\cdot\hat{\mathcal{H}}^\text{s}_k\cdot\hat{\mathcal{U}}^{\text{s}\to\text{o}}_k
    \text{.}
\end{equation}
We call this alternative form of the Hamiltonian ``sublattice form," since it treats states within the unit cell as being located on different sublattice sites. We now show that implementing the Peierls substitution with $\hat{\mathcal{H}}^\text{o}_k$ \textit{before} the unitary transformation is equivalent to taking into account both the intercell and intracell currents (we set the electric charge ${\rm q}=-1$ for simplicity),
\begin{equation}
\begin{split}
    \hat{j}^x &= \frac{\partial}{\partial A}\left(\hat{\mathcal{U}}^{\text{s}\to\text{o}\dagger}_k\hat{\mathcal{H}}^\text{s}_{k+A}\hat{\mathcal{U}}^{\text{s}\to\text{o}}_k\right) \\
    &= \frac{\partial}{\partial k}\left(\hat{\mathcal{U}}^{\text{s}\to\text{o}\dagger}_k\mathcal{H}^\text{s}_{k}\hat{\mathcal{U}}^{\text{s}\to\text{o}}_k\right) \\
    &\,\,\,\,\,\,\,-\frac{\partial}{\partial k}\hat{\mathcal{U}}^{\text{s}\to\text{o}\dagger}_k\hat{\mathcal{H}}^\text{s}_{k}\hat{\mathcal{U}}^{\text{s}\to\text{o}}_k - \hat{\mathcal{U}}^{\text{s}\to\text{o}\dagger}_k\hat{\mathcal{H}}^\text{s}_{k}\frac{\partial}{\partial k}\hat{\mathcal{U}}^{\text{s}\to\text{o}}_k \\
    &= \frac{\partial\hat{\mathcal{H}}^\text{o}_k}{\partial k} - \frac{\partial}{\partial k}\hat{\mathcal{U}}^{\text{s}\to\text{o}\dagger}_k\hat{\mathcal{U}}^{\text{s}\to\text{o}}_k\hat{\mathcal{H}}^\text{o}_{k} - \hat{\mathcal{H}}^\text{o}_{k}\hat{\mathcal{U}}^{\text{s}\to\text{o}\dagger}_k\frac{\partial}{\partial k}\hat{\mathcal{U}}^{\text{s}\to\text{o}}_k\\
    &= \frac{\partial\hat{\mathcal{H}}^\text{o}_k}{\partial k} + i\left[\hat{\mathcal{H}}^\text{o}_k\,\,,\,\, i\,\hat{\mathcal{U}}^{\text{s}\to\text{o}\dagger}_k\frac{\partial}{\partial k}\hat{\mathcal{U}}^{\text{s}\to\text{o}}_k\right] \\
    &\equiv \hat{v}^x_\text{inter} + \hat{v}^x_\text{intra}
    \text{.}\nonumber
\end{split}
\end{equation}
We therefore recognize the identity $i\,\hat{\mathcal{U}}^{\text{s}\to\text{o}\dagger}_k\frac{\partial}{\partial k}\hat{\mathcal{U}}^{\text{s}\to\text{o}}_k \equiv \hat{\tau}$, or in other words, $\hat{\mathcal{U}}^{\text{s}\to\text{o}}_k \equiv e^{-ik\hat{\tau}}$. In the case at hand, we have, (up to an overall phase representing a shift in origin)
\begin{equation}
    \hat{\mathcal{U}}^{\text{s}\to\text{o}}_k =
    \begin{pmatrix}
    e^{i k/4} & 0 & 0 & 0 \\
    0 & e^{-i k/4} & 0 & 0 \\
    0 & 0 & e^{-i k/4} & 0 \\
    0 & 0 & 0 & e^{i k/4}
    \end{pmatrix}
\end{equation}
\begin{equation}
    \implies \hat{\tau} =
    \begin{pmatrix}
    -\frac{1}{4} & 0 & 0 & 0 \\
    0 & \frac{1}{4} & 0 & 0 \\
    0 & 0 & \frac{1}{4} & 0 \\
    0 & 0 & 0 & -\frac{1}{4}
    \end{pmatrix}
    \text{.}
\end{equation}
That is, the position operator acquires an intracell part encoding the fact that two of the sublattices are displaced in the $x$-direction by half the width of the unit cell. Therefore, in what follows, we use the following form for the Peierls-substituted Hamiltonian,
\begin{equation}
    \hat{\mathcal{H}}^\text{s}_k = -
    \begin{pmatrix}
        \Delta & g_A(k) & 0 & t_y \\
        g_A^*(k) & -\Delta & t_y & 0 \\
        0 & t_y & -\Delta & g_A^*(k) \\
        t_y & 0 & g_A(k) & \Delta
    \end{pmatrix}
    \text{,}
\end{equation}
\begin{equation}
    g_A(k) = e^{-i k/2}\left[(t+\delta)e^{i(k-A)/2}+(t-\delta)e^{-i(k-A)/2}\right]
    \text{,}
\end{equation}
which has \textit{both} the periodicity $k\to k+2\pi$ \textit{and} a dependence on $A$ that encodes both intracell and intercell currents.

We can then define the current operators in the $x$-direction in terms of derivatives with respect to $A$,
\begin{equation}
    \hat{v}^{x} \equiv \hat{h}^{x} = \left.\hat{\mathcal{U}}^\dagger_k\cdot\frac{\partial\hat{\mathcal{H}}^A_k}{\partial A}\cdot\hat{\mathcal{U}}_k\right\vert_{A=0}
    \text{,}\label{vx}
\end{equation}
\begin{equation}
    \hat{h}^{xx} = \left.\hat{\mathcal{U}}^\dagger_k\cdot\frac{\partial^2\hat{\mathcal{H}}^A_k}{\partial A^2}\cdot\hat{\mathcal{U}}_k\right\vert_{A=0}
    \text{,}\label{hxx}
\end{equation}
\begin{equation}
    \hat{h}^{xxx} = \left.\hat{\mathcal{U}}^\dagger_k\cdot\frac{\partial^3\hat{\mathcal{H}}^A_k}{\partial A^3}\cdot\hat{\mathcal{U}}_k\right\vert_{A=0}
    \text{,}\label{hxxx}
\end{equation}
where $\hat{\mathcal{U}}_k$ is some (non-unique) unitary transformation that diagonalizes $\hat{\mathcal{H}}^{A=0}_k$. As for the current operators in the $y$-direction, since there is no periodicity in $y$, we use,
\begin{equation}
    \hat{v}^y = i\,\hat{\mathcal{U}}^\dagger_k\cdot\left[\hat{\mathcal{H}}^\text{o}_k\,,\,\hat{r}^y\right]\cdot\hat{\mathcal{U}}_k
    \text{,}
\end{equation}
with $\hat{r}^y$ defined just as in the molecule case (Eq. (\ref{r operators})). For simplicity, we set the width of the chain in the $y$-direction to be $1/2$, to match up with the half-unit-cell width in the $x$-direction of a single decoupled molecule. (That is, in comparing with Eq. (\ref{r operators}), we will set $R_x=R_y=1/4$.) One can then formally compute higher-order currents in the $y$-direction using the covariant derivative, where in this case there is no partial derivative, and so the covariant derivative reduces to a commutator with the Berry connection $\hat{\mathcal{A}}^y \equiv \hat{r}^y$,
\begin{equation}
    \hat{h}^{yy} = -i\left[ \hat{\mathcal{A}}^y \,,\,\hat{v}^y\right]
    \text{,}
\end{equation}
and so on for all other current operators $\hat{h}^{\alpha_1\hdots\alpha_n}$.

\subsection{Numerical results}

Figures \ref{fig:qLIN} and \ref{fig:qSHG} show numerical results for the spatially dispersive corrections to the linear conductivity and SHG conductivity, respectively. We benchmark our methodology in two ways. First, we compare the conductivities of the chain in the $t_x'\to 0$ limit to the conductivities calculated for a single rectangular molecule. Second, for the linear conductivity, we also benchmark against established results \cite{Malashevich2010, Ahn2022} for the spatially dispersive linear conductivity calculated by expanding the Kubo formula to first order in $\mathbf{q}$. Figures \ref{fig:qLIN} and \ref{fig:qSHG} show agreement in all cases. In our plots, we introduce a phenomenological relaxation parameter $\eta$ by taking $\omega \to \omega + i\eta$, with $\eta=0.01$. We set $t_x=1$, without loss of generality, and in the case for the linear conductivity, we set $\Delta=0.1$ in both the chain and the molecule to statically break inversion (and to therefore allow a nonzero spatially dispersive correction). We also always assume the \textit{lowest three bands} in this model are filled,
\begin{equation}
    \hat{\rho}^{(0)} =
    \begin{pmatrix}
        1 & 0 & 0 & 0 \\
        0 & 1 & 0 & 0 \\
        0 & 0 & 1 & 0 \\
        0 & 0 & 0 & 0
    \end{pmatrix}
    \text{,}
\end{equation}
because the spatially dispersive correction to SHG depends on a peculiar combination of the filling factors (e.g. for the molecule, it is proportional to $f_1 - f_2 - f_3 + f_4$).

As a note on symmetries, the molecular chain we consider is inversion symmetric when \textit{either} $\Delta=0$ \textit{or} $\delta=0$, where we define $\delta = t_x - t_x'$. This is only evident from the ``sublattice form" $\hat{\mathcal{H}}_k^\text{s}$ of the Hamiltonian in Eq. (\ref{sublattice form}). When $\delta=0$, the system has an inversion center on a vertical bond, represented by the following matrix transformation,
\begin{equation}
    \hat{\mathcal{H}}_k^\text{s} \longrightarrow \hat{\mathcal{H}}_{-k}^\text{s}
    \text{.}
\end{equation}
Additionally, the system has an inversion center at the \textit{midpoint} of the unit cell whenever $\Delta=0$, represented by,
\begin{equation}
    \hat{\mathcal{H}}_k^\text{s} \longrightarrow 
    \begin{pmatrix}
        0 & 0 & 1 & 0 \\
        0 & 0 & 0 & 1 \\
        1 & 0 & 0 & 0 \\
        0 & 1 & 0 & 0
    \end{pmatrix}
    \cdot\hat{\mathcal{H}}_{-k}^\text{s}\cdot
    \begin{pmatrix}
        0 & 0 & 1 & 0 \\
        0 & 0 & 0 & 1 \\
        1 & 0 & 0 & 0 \\
        0 & 1 & 0 & 0
    \end{pmatrix}
    \text{.}
\end{equation}
So inversion is only broken when \textit{both} $\delta\neq 0$ and $\Delta\neq 0$.

\begin{figure}[t]
    \centering
    \includegraphics[width=\linewidth]{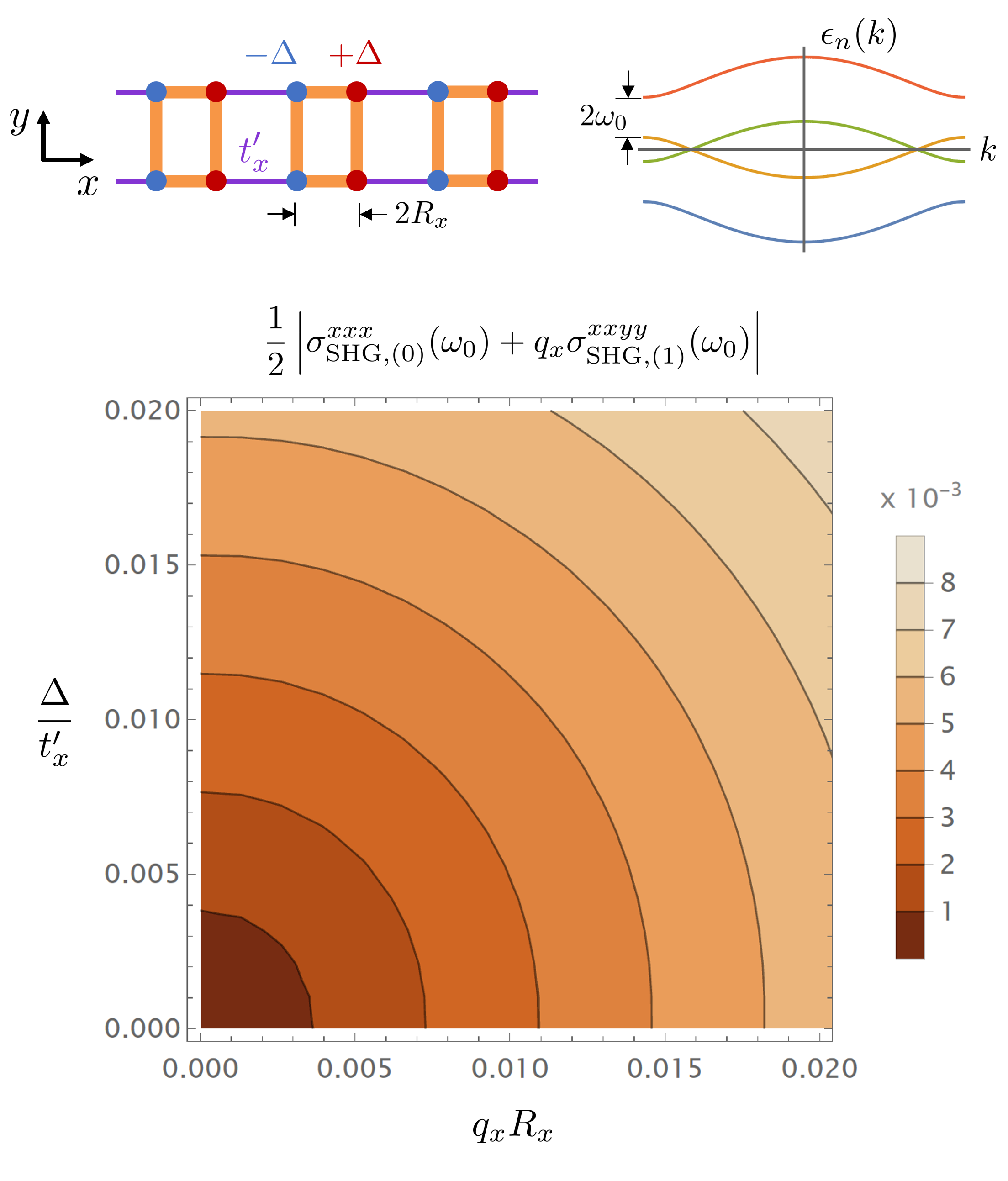}
    \caption{Magnitude of the \textit{total} SHG response computed up to electric quadrupole order resulting from off-axis linearly polarized light incident on the molecular chain model, evaluated at $\omega_0 = \sqrt{(t_x-t_x')^2+\Delta^2}$ (half the minimum gap between the filled and empty bands). Light polarized in $x$ (chain direction) leads to a conventional SHG response when inversion is broken ($\Delta \neq 0$) while light polarized in $y$ leads to a spatially dispersive correction ($q \neq 0$). Here, $t_x-t_x'=0.5$ and $t_y=0.8$.}
    \label{fig:mag total}
\end{figure}

Given these considerations, we can directly compare SHG resulting from static inversion breaking with SHG resulting from a spatially dispersive correction. Figure \ref{fig:mag total} plots the magnitude of the \textit{total} SHG response resulting from diagonally polarized light at a frequency of half the minimum gap in this model. We choose diagonal polarization because $x$-polarized light triggers a \textit{uniform} SHG response ($\sigma_{\text{SHG},(0)}^{xxx}$) when $\Delta$ is nonzero, while $y$-polarized light triggers a spatially dispersive correction ($\sigma_{\text{SHG},(1)}^{xxyy}$) at first order in $q_x$. In that plot, we choose a window such that the contours are roughly circular in shape, and find a pair of dimensionless parameters in which the two axes have roughly the same scale. We find these relevant dimensionless parameters to be $\Delta/t_x'$ and $q_x R_x$. That is, the effects of spatial dispersion on the length scale of half the width of an individual molecule create a similar magnitude of SHG to that of a static inversion breaking parameter on the energy scale of the inter-cell hopping. This heurstic has intriguing implications for real systems where spatially dispersive effects may be present. For instance, moir\'e systems famously create extremely inflated artificial lattice constants, in which case $q R$ is very large (where $R$ is interpreted generally as some coherence length for localized electronic states). So long as the hopping amplitude between superlattice sites is not too small, the spatially dispersive correction to SHG may be on the same order as a typical static inversion-breaking mechanisms for which SHG is often used as a probe. This has important ramifications for using SHG to detect spontaneous breaking of inversion symmetry, since many systems of recent interest may have sufficiently large coherence length scales $R$ to allow spatially dispersive effects to dominate such a response.

\section{Outlook}

We have introduced a new scheme for calculating spatially dispersive corrections to nonlinear optical responses in velocity gauge. Reviewing from a diagrammatic standpoint the subtle cancellations that make up the sum rules guaranteeing equivalence between the length and velocity gauges, we compactly write velocity gauge expressions for nonlinear optical conductivities and introduce spatially dispersive corrections as an additional vertex rule. Though we only treat the electric quadrupole contribution here, we expect the formalism can be straightforwardly extended to include the magnetic dipole contribution, as well as higher-multipole corrections as desired. We expect spatially dispersive corrections to be important for understanding anomalously strong bulk second-order responses in centrosymmetric materials where the response is typically expected to vanish.

Our results put forward a new tool for using nonlinear optics to study quantum materials beyond electric dipole order. We expect spatially dispersive corrections to optical responses to reveal information about intrinsic nonlocality in material systems, since they should be strong when the electric field varies appreciably on the length scale of the electronic states being coupled by optical transitions. A number of materials of recent interest fall into this category, including Moir\'e materials \cite{Koshino2018, Kang2018} and systems whose bands exhibit nontrivial quantum geometry \cite{Brouder2007, Marzari2012, Po2018} for which maximally localized Wannier representations still feature coherence over length scales of many unit cells. Searching for measurable spatially dispersive corrections in these systems and linking them to quantum geometry is a promising direction for future work.

\begin{acknowledgments}
The authors thank M. Claassen, C. De Beule, Z. Addison, and L. Wu for insightful discusssions. S.G. was supported by an NSF Graduate Research Fellowship. This work was supported by the Department of Energy under Grant DE-FG02-84ER45118.
\end{acknowledgments}

\newpage

\bibliography{references}

\onecolumngrid
\pagebreak

\end{document}